\documentclass[11pt]{article} 

\usepackage{amsmath,amsthm,latexsym,amssymb,amsfonts,epsfig}

\usepackage{changepage}

\newtheorem*{Theorem}{Theorem}

\newcommand{\be}{\begin{equation}}
\newcommand{\ee}{\end{equation}}
\newcommand{\ba}{\begin{eqnarray}}
\newcommand{\ea}{\end{eqnarray}}

\title{{\sf Reduced Phase Space Approach to the}\\
 {\sf $U(1)^3$ model for Euclidean Quantum Gravity}} 
\author{
{\sf S. Bakhoda}$^{1,2}$\thanks{{\sf 
s\_bakhoda@sbu.ac.ir, sepideh.bakhoda@gravity.fau.de}},
{\sf T. Thiemann}$^2$\thanks{{\sf 
thomas.thiemann@gravity.fau.de}}\\
\\
{\sf $^1$ Department of Physics, Shahid Beheshti University,}
{\sf Tehran, Iran}\\
{\sf $^2$ Institute for Quantum Gravity, FAU Erlangen -- N\"urnberg,}\\
{\sf Staudtstr. 7, 91058 Erlangen, Germany}\\
}
\date{{\small\sf \today}}

\makeatletter
\@addtoreset{equation}{section}
\makeatother

\begin{document} 

\maketitle

{\sf

\begin{abstract}
If one replaces the constraints of the Ashtekar-Barbero $SU(2)$ gauge theory 
formulation of Euclidean gravity by their $U(1)^3$ version, one arrives at a 
consistent model which captures significant structure of its $SU(2)$ version.
In particular, it displays a non trivial realisation of the hypersurface 
deformation algebra which makes it an interesting testing ground for (Euclidean)
quantum gravity as has been emphasised in a recent series of papers due 
to Varadarajan et al. 

In this paper we consider a reduced phase space approach to this model. This 
is especially attractive because, after a canonical transformation, the constraints 
are at most {\it linear} in the momenta. In suitable gauges, it is therefore 
possible to find a closed and explicit formula for the physical Hamiltonian 
which depends only on the physical observables. 

The corresponding 
reduced phase space quantisation can be confronted with the constraint 
quantisation due to Varadarajan et al to gain further insights into the 
quantum realisation of the hypersurface deformation algebra.        
\end{abstract}

\section{Introduction}
\label{s1}

Using real-valued SU(2) gauge theory variables \cite{1} one can write the Hamiltonian constraint of Lorentzian vacuum General Relativity (GR) as a sum of three terms.  
The first term can be recognised as the Hamiltonian constraint for Euclidean vacuum
GR, the second term corrects for Lorentzian signature and the third is the 
cosmological constant term. The second term 
can, in turn, be expressed as a multiple Poisson bracket between the cosmological 
constant term and the first term \cite{2}.  Following the quantisation rule 
that Poisson brackets be replaced by commutators, the Lorentzian Hamiltonian quantum constraint can thus be written as a linear combination of the Euclidean 
Hamiltonian quantum constraint, the quantum cosmological constant term, and multiple 
commutators thereof \cite{2}. Since the quantum cosmological constant term is 
essentially the volume operator\footnote{By the 4-volume resp. 3-volume term we mean $|\det(g)|^{1/2}$ and $\det(q)^{1/2}$ respectively. In any ADM slicing we have the identity $|\det(g)|^{1/2}=N \det(q)^{1/2}$ with the lapse function $N$. Since the constraints concern single moments of the slicing time parameter, the relation between the two terms is really that simple.} \cite{3} which in Loop Quantum Gravity (LQG)
\cite{4} is under complete control, the quantisation of Lorentzian vacuum 
gravity is reduced essentially to the quantisation of the 
Euclidean Hamiltonian constraint 
which has been intensively studied \cite{5}. 

A critical measure for the success of that quantisation is a faithful representation of the hypersurface 
deformation algebra which is formed by the Euclidean Hamiltonian constraint 
together with the spatial diffeomorphism constraint. The quantisation 
of \cite{5} displays a closed quantum constraint algebra, however, while 
the commutator of two Euclidean Hamiltonian constraints is a linear combination 
of spatial diffeomorphism constraints, the quantum coefficients of that linear combination are not quantisations of the coefficients of the corresponding classical linear
combination, i.e. the ``quantum structure functions" differ from the classical 
ones. 

In order to find clues for how to improve on this, one can study the same problem in the simpler $U(1)^3$ model \cite{6} which recently was done in the series of works \cite{7}. The $U(1)^3$ model is defined by replacing the SU(2) Gauss constraint
and the SU(2) version of the Euclidean Hamiltonian constraint by their corresponding
$U(1)^3$ counterparts while the symplectic structure and the spatial diffeomorphism constraints remain unchanged. Thus essentially one replaces the SU(2) covariant derivative and curvature respectively by their $U(1)^3$ versions. The resulting
spatial diffeomorphism and Hamiltonian constraints form an algebra isomorphic to its SU(2) version with the same structure functions, i.e. they form a 
hypersurface deformation algebra. This makes it an ideal testing ground for the actual SU(2) problem under investigation, in particular, the model has two propagating degrees of freedom.  

In this paper, we propose to quantise the $U(1)^3$ model using the reduced phase 
space approach \cite{8} rather than by operator constraint methods \cite{7}.
The advantage of the reduced phase space approach is that one can work directly with the physical Hilbert space, the physical observables, and the physical dynamics
which in the operator constraint approach have to be determined after representing
the constraints on an unphysical, kinematical Hilbert space. The disadvantage of the reduced phase space approach is that generically it involves implicit functions, inversions of phase space-dependent differential operators, and square roots which are difficult to quantise in practice \cite{9}. This problem can be avoided in the presence of suitable matter \cite{10} but in vacuum it is generic. 

What makes the reduced phase space approach a practical possibility for the vacuum 
$U(1)^3$ model is the fact that its constraints are {\it at most linear} in the 
$U(1)^3$ connection while in the SU(2) case the connection dependence is 
{\it quadratic}. Considering the connections as momentum variables after a 
trivial canonical transformation, this means that no square roots have to be 
taken to solve the constraints.
  Even more, the explicit solution of the constraints is not even necessary in order to construct the physical Hamiltonian corresponding to a choice of gauge fixing in terms of the corresponding physical
degrees of freedom. Depending on the choice of gauge fixing, 
that Hamiltonian generically still involves inversions of phase space-dependent differential operators for which we can give explicit perturbative expressions which however will be difficult to quantise. However, we find preferred gauge fixings which avoid this problem. The resulting physical Hamiltonians are spatially non-local, however, they can be written explicitly
and non-perturbatively and is not hopeless to be quantisable using LQG methods. 
The resulting equations of motion for the 
physical observables can be worked out explicitly. For the electric 
field\footnote{Throughout this paper, by the electric field, we mean a densitised triad.} these do not involve the connection and are non-polynomial differential equations of second order in time and involve higher order integro-differential operators. 
By passing to suitable variables one can write these as fourth-order polynomial equations without integral operators at the price of introducing higher-order spatial derivatives. The equations of motion 
for the connection are linear in the connection and of first-order in time 
and thus can in principle be easily
integrated once the solution for the electric field dynamics has been found. 
The physical dynamics is such that the two 
physical degrees of freedom are self-interacting and propagating.  

We consider it quite remarkable that a model so closely related to vacuum GR 
can be cast into a rather manageable form while the symplectic reduction of all constraints has already been carried out. 
This paper serves as a proof of principle that such a possibility exists. In future publications, we plan to further optimise the 
choice of gauge fixing in order to simplify the physical Hamiltonian as much as possible and then to find Hilbert space representations thereof. 
We hope that such a complementary approach to the developments of \cite{7}
will shed further light on the actual problem of quantisation of the Euclidean 
SU(2) Hamiltonian constraint. Another direction of generalisation is the consideration of non-trivial spatial topology, in this paper we restrict to the topology of $\mathbb{R}^3$ supplemented by asymptotically flat boundary conditions. The details of these boundary conditions can be found in \cite{11} 
which is based on \cite{12,13}.
\\
\\
The architecture of this paper is as follows:\\
\\
In section 2 we briefly review the classical formulation of the $U(1)^3$ model which mainly serves to fix our notation. We also summarise the asymptotically flat boundary conditions.

In section 3 we prove a simple lemma concerning the construction of the physical Hamiltonian defined by a choice of gauge fixings for a first-class constrained system with constraints which are at most linear in the momenta.
The most important conclusion from that lemma is that the constraints do not 
have to be solved explicitly in order to find the physical Hamiltonian. 
This result may sound trivial for systems with a finite number of degrees of freedom
where only a matrix has to be inverted to solve the constraints 
but in field theory, the result is of major practical importance because the 
matrix is generically replaced by a system of differential operators with coefficients 
which depend on the configuration degrees of freedom, thus are not known as 
explicit functions of the spatial variables and are thus only implicitly 
invertible.  

In section 4 we apply our lemma to the $U(1)^3$ model for various choices of gauge fixings to construct the physical Hamiltonian. The degree of spatial 
non-locality and non-polynomiality 
critically depend on that choice. While these choices are classically equivalent
(related by a gauge transformation), with regard 
to quantisation certain choices seem to be preferred. We give an explicit example for which the physical Hamiltonian displays a rather manageable form,
find the physical equations of motion and discuss its properties.

In section 5 we conclude and give an outlook to future research.\\   
\\
This paper is the first in a series of three. In our companion papers 
\cite{11,15} we study more systematically the asymptotically flat boundary 
conditions that we chose throughout this paper as we need them in various 
places in order to invert differential operators (i.e. construct Green 
functions). We also complement the present canonical analysis by 
a covariant framework.

\section{Review of the $U(1)^3$ model for Euclidean vacuum GR}
\label{s2}               

We consider first the connection formulation of General Relativity \cite{1} in terms 
of $su(2)$-valued gauge theory variables $(A_a^j,E^a_j)$ where $a,b,c,..\in \{x,y,z\}$
are spatial tensor indices and $j,k,l,..\in \{1,2,3\}$ are $su(2)$ Lie algebra 
indices that are moved with the Kronecker $\delta^i_j$ and so the index position of $j,k,l,\dots$  is irrelevant. These variables are canonically conjugate in the sense that the non 
vanishing Poisson brackets are  
\be \label{2.0}
\{E^a_j(x),A_b^k(y)\}=\kappa\; \delta^a_b\; \delta_j^k\; \delta(x,y)
\ee
where $\hbar\kappa=\ell_P^2$ is the Planck area. Geometrically, $E$ is a density 
one valued triad for the geometry intrinsic to the spatial hypersurfaces 
and $A$ differs from the spin connection $\Gamma$ of $E$ by a term 
linear in their extrinsic curvature.  
    
In what follows we assume for simplicity that the spacetime topology is that of 
$\mathbb{R}^4$ and 
that asymptotically flat boundary conditions are imposed in the connection  
formulation of GR \cite{11,12,13}. This in particular implies that 
\be \label{2.1}
E^a_j-\delta^a_j\to F^a_j(n)/r+O(1/r^2),\;
A_a^j \to G_a^j(n)/r^2+O(1/r^3)
\ee 
where $r$ is an asymptotic radial variable $r^2=x^2+y^2+z^2$, $n^a=x^a/r$ is an 
asymptotic angular variable and the functions $F,G$ on the asymptotic $S^2$ are 
even and odd respectively $F(-n)=F(n), G(-n)=-G(n)$. The meaning of $\delta^a_j$ 
is that it is non-vanishing only for the pairs $(a,j)=(x,1), (y,2), (z,3)$ and then 
takes the value one. We will simply copy these 
boundary conditions as well as the symplectic structure (\ref{2.0})
into the $U(1)^3$ truncation of (Euclidean) GR, the motivation
being the importance of asymptotically flat spacetimes in 
Lorentzian GR which we thus wish to probe as closely as possible with the $U(1)^3$ 
model. Further details including the asymptotic symmetry (ADM) charges can be found 
in \cite{11}. 

The first class, density one vacuum constraints of Euclidean GR with vanishing 
cosmological constant
can be written in these variables 
as
\be \label{2.2}
C_j=D_a E^a_j,\;
C_a=F_{ab}^j E^b_j,\;
C_0=F_{ab}^j \delta^{jk}\epsilon_{klm}\; E^a_l\; E^b_m \;|\det(E)|^{-1/2}
\ee
where $D_a$ is the $su(2)$ gauge covariant derivative and $F=2(dA+A\wedge A)$ 
the curvature of $A$. The constraints are respectively referred to as the 
Gauss-, vector- and scalar constraint respectively. 

Lorentzian GR is now obtained by adding to $C_0$ a 
term proportional to \cite{2}
\be \label{2.3}
{\rm Tr}(\{\{\int d^3y C_0(y),V\},A(x)\}
\wedge \{\{\int d^3y C_0(y),V\},A(x)\}
\wedge \{V,A(x)\}),\; V=\int d^3y |\det(E)|^{1/2}
\ee
where $V$ is the total volume of the spatial hypersurfaces. In this paper 
we will only be concerned with the Euclidean piece which already leads to 
the hypersurcae deformation algebra.

The truncation of the model is obtained by writing 
$A=\kappa \tilde{A}$ so that $E,\tilde{A}$ are strictly conjugate without a factor 
of $\kappa$ and by expanding (\ref{2.2}) to first non-vanishing order in 
$\kappa$ which can be considered as a weak coupling limit with respect to 
Newtons constant $\kappa$ \cite{6}. This simply means that (\ref{2.2}) is 
replaced by
\be \label{2.4}
C_j=\partial_a E^a_j,\;
C_a=F_{ab}^j E^b_j,\;
C_0=F_{ab}^j \delta^{jk}\epsilon_{klm}\; E^a_l\; E^b_m \;|\det(E)|^{-1/2}
\ee
where $\partial$ is the spatial derivative and $F=2 dA$. This already shows 
that the model depends on $A$ only linearly. Remarkably, as one can check,
the constraint algebras defined by (\ref{2.2}) and (\ref{2.4}) induced by 
(\ref{2.0}) and restricted to the spacetime diffeomorphism sector are 
{\it identical} and form the hypersurface deformation algebra
\be \label{2.5}
\{C(u,f),C(v,g)\}=-\kappa C([u,v]+q^{-1}[f\; dg-g\; df], u[g]-v[f])
\ee
with 
\be \label{2.6}
C(u,f)=\int\; d^3x\;[u^a C_a+f C_0];\;\; [q^{-1}]^{ab}=E^a_j E^b_k\delta^{jk}
|\det(E)|^{-1} 
\ee
Also $\{C(\lambda),C(U,f)\}=0$ is the same as as in the $su(2)$ theory where 
$C(\lambda)=\int d^3x \lambda^j C_j$. The only difference is that in in $su(2)$ 
we have $\{C(\lambda),C(\lambda')\}=-\kappa C([\lambda,\lambda'])\not\equiv 0$ 
while in $u(1)^3$
we have $\{C(\lambda),C(\lambda')\}\equiv 0$. 

For our analysis it will prove to be convenient to introduce the following 
density one valued quadratic combinations
\be \label{2.7} 
H_a^j=\frac{1}{2} \epsilon_{abc} \; \epsilon^{jkl} E^b_k E^c_l
=\det(E)\; (E^{-1})_a^j 
\ee
which allows us to write vector and scalar constraint in an equivalent but 
unified density two valued form 
\be \label{2.8}
\tilde{C}_j=\epsilon_{jkl} \delta^{km} B^a_m H_a^l,\; 
\tilde{C}_0=B^a_j H_a^j
\ee
where 
\be \label{2.9}
B^a_j=\epsilon^{abc} \partial_b A_c^k \delta_{kj}
\ee
defines the magnetic field\footnote{Throughout this paper, by the magnetic field, we mean the dual of the derivative of the spin connection.} of $A$. To arrive at this form, it was implicitly 
assumed as usual that the triad is nowhere degenerate. Following 
\cite{14}, it is now apparently possible to solve the constraints algebraically: Let
\be \label{2.10}
B^a_j=c^{jk} \delta_{kl} E^a_l 
\ee
then the spacetime constraints simply demand that $c=c^T,\; {\rm Tr}(c)=0$.
However, this is not the case as the Bianchi identity and the Gauss constraint
in addition implies 
\be \label{2.11}
E^a_j \partial_a c^{kj}=0
\ee
which presents a system of three coupled PDE's of first order which is no longer 
solvable algebraically. Thus, in order to proceed the gauge fixing analysis of the 
subsequent section is unavoidable.

We note that the linearity of the constraints in $A$ implies that the gauge 
transformations generated by the constraints on $E^a_j$ have a purely geometric
interpretation: Obviously, $E^a_j$ is Gauss invariant and transforms as a 
vector density under spatial diffeomorphisms. A scalar gauge transformation has the 
effect (using the density one version)
\be \label{2.11}
\delta_f E^a_j:=\{C_0(u=0,f),E^a_j\}=-\epsilon^{abc} \partial_b [f e^j_c]
\ee
where $e^j_c$ is the density zero co-triad. It follows that the scalar constraint
shifts the divergence free piece of the electric field which is in particular
consistent with the Gauss constraint. 

This suggests to consider the canonical transformation $(A,E)\mapsto (E,-A)$ and 
to consider $-A$ as the momentum conjugate to $E$. We call this the $(A,E)$ 
formulation. We consider also a dual $(f,B)$ formulation as follows: It is 
consistent with the boundary conditions to solve the Gauss constraint in the 
form $E^a_j=\delta^a_j +\epsilon^{abc} \partial_b f_c^j$. When plugging this into 
the symplectic potential we obtain upon integration by parts 
\be \label{2.12}
\theta
=\int\; d^3x\; E^a_j\delta A_a^j
=\delta[\int\; d^3x\; E^a_j A_a^j]
-\int\; d^3x\; A_a^j \delta [E^a_j-\delta^a_j]
=\delta[\int\; d^3x\; E^a_j A_a^j]
-\int\; d^3x\; B^a_j \delta f_a^j
\ee
which displays $-B^a_j$ as conjugate to $f_a^j$. Note that the roles of $(f,-B)$ have 
changed as compared to $(A,E)$: While the Gauss constraint only depends on $E$
and just transforms $A$, there is now no longer a Gauss constraint but instead the 
Bianchi identity $\partial_a B^a_j=0$ which, when viewing $B$ as an elementary
rather than a derived field, plays the role of a ``dual Gauss constraint''.
Likewise, the Bianchi identity just depends on $B$ and only 
transforms $f_a^j$ and shifts its curl free 
piece. The $(B,f)$ formulation thus has the advantage that all constraints 
(dual Gauss constraint, spacetime constraints) are linear in $B$ and the 
spacetime constraints contain the momentum $B$ conjugate to $f$ only algebraically.

\section{Physical Hamiltonian for first class constrained systems with certain
linearity properties}
\label{s3}

In this section, we state and prove two simple results on the reduced dynamics 
of constrained systems which have a certain linear structure in their constraints 
with respect to the momenta. This linear structure will be tailored to the 
appearance of the constraints of the $U(1)^3$ model displayed in the previous 
section. Hence the subsequent results  will be applicable to the $U(1)^3$ model 
both in its $(A,E)$ and its $(B,f)$ description.
In what follows, as customary, we distinguish between constraint surface (constraints vanish) and reduced phase space (constraint surface modulo gauge transformation) or (locally) equivalently the gauge cut (constraints and gauge conditions vanish).

\subsection{The (A,E) description}
\label{s3.1}

We have seen that the $U(1)^3$ model has seven first class constraints: Three Gauss 
constraints $C_j(x),\; j=1,2,3$ for each spatial point $x$ which do not depend on the 
connection $A_a^j$, three density weight two 
vector constraints $\tilde{C}_j(x)$ linear in the connection and one density weight 
two Hamiltonian 
constraint $\tilde{C}_0(x)$ also linear in the connection. We integrate the 
constraints 
with respect to a smooth orthonormal basis of test functions $b^\alpha$ 
on $L_2(d^3x,\;\sigma)$ where $\alpha$ 
takes values in a suitable index set
thus obtaining 
$C_j^\alpha=\int\; d^3x\; C_j(x)\; b^\alpha(x)$ etc. We denote the 
$C_j^\alpha$ collectively by $C_A$ and the $\tilde{C}_\mu^\alpha,\; \mu=0,..,3$
by $C_I$ for suitable ranges of the indices $A,I$. We can integrate 
the canonical pairs $(A_a^j, E^a_j)$ with respect to the same test functions 
and subdivide them as $w=(u^A,v_A),\; z=(x^I,y_I),\; 
r=(q^\mathfrak{a},p_\mathfrak{a})$. These are still conjugate pairs and in 
particular all variables from $w$ have vanishing Poisson brackets with all variables 
from $z,r$ and all variables from $z$ have vanishing Poisson brackets with the 
variables from $r$. Then using the completeness relation for the 
basis $b^\alpha$ we can write the constraints in the form 
\be \label{3.1}
C_A=C_A(u,x,q),\; C_I=M_I\;^J(u,x,q) \; y_J+N_I\;^A(u,x,q)\; v_A+h_I(u,x,q,p)
\ee
In our case $C_A$ depends linearly on $u,x,q$ and $h_I$ depends linearly on $p$
but we will not need to use this.
Moreover, $M_I\;^J,\;N_I\;^A,\; h_I$ are 
homogeneous and quadratic in $u,x,q$ but we will not need to use this either.  
What we need is that the subdivision of the canonical pairs into 
groups is done in such a way 
that the ``matrices'' $\sigma_{AB}:=\{C_A, v_B\}$ and $M_I\;^J$ are non-singular
and this guides the above subdivision of canonical pairs. This means that 
$G_A:=v_A=0$ is a suitable gauge fixing condition for $C_A$ and $G^I:=F^I(x)-\tau^I$ 
for $C_I$ where $\tau^I$ are constants on the phase space but possibly 
functions of physical time $\tau$ and 
$\Delta_J\;^I:=\{y_J,F^I\}$ is non-singular. Indeed, recall \cite{8} 
that depending on whether
the spatial manifold $\sigma$ has a boundary or not, a time dependence of 
$\tau$ is not necessary or necessary respectively in order that the physical
dynamics be non-trivial due to the different boundary conditions on the 
canonical variables in these two cases. Our assumptions imply, 
by the implicit function 
theorem,  that $C_A=0$ can locally be solved for 
$u^A=g^A(x,q)$ and $G^I=0$ for $x^I=k^I(\tau)$ for suitable functions $g^A, k^I$. 

We thus declare 
$r$ as our physical degrees of freedom. 
Note that in field theory the statement that infinite dimensional matrices are 
non singular is to be taken with care: If for instance 
differential operators are involved
then an inverse of say $M_I\;^J$ only exists if one specifies a suitable function 
space possibly accompanied with boundary conditions on them such that otherwise 
unspecified constants of integration are uniquely fixed. 

The first class Hamiltonian is given by 
\be \label{3.2}
H(\lambda,\Lambda)=\lambda^A \; C_A+ \Lambda^I \; C_I
\ee
where $\lambda^A$ are the Lagrange multipliers of the Gauss constraint integrated 
against the basis and likewise $\Lambda^I$ are the density weight minus one 
lapse and shift functions integrated against that basis, in particular they are 
phase space independent. We collectively denote them by 
$l=(\lambda,\Lambda)$. The stability of the gauge 
conditions under gauge transformations fixes the Lagrange multipliers 
\be \label{3.3}
\dot{G}_A=\{H,G_A\}=\lambda^B\;\sigma_{BA}+\Lambda^I \{C_I,G_A\}=0;\;\;    
\dot{G}^I=\{H,G^I\}=\Lambda^J\; M_J\;^K\; \Delta_K\;^I=\dot{\tau}^I
\ee
which has the explicit solution
\be \label{3.4}
\lambda_0^A=-\Lambda_0^I\; \{C_I,G_B\}\; (\sigma^{-1})^{BA}+\kappa^A,\;
\Lambda_0^J M_J\;^I=(\Delta^{-1})_J^I \; \dot{\tau}^J+\kappa^I=:\delta^I,\;
\Lambda_0^I =(M^{-1})_J\;^I\;\delta^J+\hat{\kappa}^I
\ee
Here we assumed that both $\sigma,\Delta$ have an unambiguous inverse 
$\sigma^{-1}, \Delta^{-1}$ on a suitable space of functions and we allow that
the space of Lagrange multiplier 
functions considered is larger and contains a kernel of $\sigma, \Delta$
leading to the ``integration constants'' $\kappa^A,\kappa^I$. By 
construction, $\kappa^A$ can only depend on $u,x,q$ and $\kappa^I$ only 
on $x$. In the $U(1)^3$ model, $\kappa^A$ is in fact phase space independent. 
As far as $M,M^{-1}$ is concerned, we allow for a similar kernel function 
$\hat{\kappa}^I$ which in general may depend on $u,x,q$. 

The ambiguities 
$\kappa^A, \kappa^I,\hat{\kappa}^I$ are supposed to be fixed by the boundary
conditions and we will assume that they imply in particular $\kappa^A=0$.    
We emphasise that $l_0=(\lambda_0, \Lambda_0)$ are phase space-dependent and are 
defined by (\ref{3.4}) on the whole phase space and not only on the 
constraint hypersurface $C_A=C_I=0$ or the reduced phase space 
$C_A=G_A=C_I=G^I=0$. However, note that $\Lambda_0$ does not depend on the 
momenta $v,y,p$ and $\Lambda_0^I M_J\;^I=\delta^I$ depends only on $x$. 

Let now $\mathcal{F}=\mathcal{F}(r)$ be a function on the reduced phase space. Its evolution is 
the gauge motion defined by $H$ restricted to the reduced phase space 
that is 
\be \label{3.5}
\dot{\mathcal{F}}=\{H,\mathcal{F}\}_{C=0,G=0,l=l_0}
\ee
If it exists, the physical Hamiltonian $h$ is a function $h=h(r,\tau,\dot{\tau})$ 
such that 
$\dot{\mathcal{F}}=\{h,\mathcal{F}\}$.

Abstracting from the concrete $U(1)^3$ model we can now state the following general 
result:
\begin{Theorem} \label{th3.1} ~\\
Let $C_A, C_I, G_A, G^I$ be as above. Then 
\be \label{3.6}
h=(\Lambda_0^I h_I)_{C_A=0,G^I=0}
\ee
\end{Theorem}
This means that we do not need to solve $C_I=0$ for $y_I$, we only need to 
solve for $\Lambda^I$ as in (\ref{3.4}) and not $\lambda^A$ and only at $C_A=G^I=0$ 
which only is 
a restriction on the configuration degrees of freedom. Given the flexibility 
in choosing the gauge condition $G^I$ this makes it conceivable that one can
arrive at an explicit expression for $h$.\\
Proof:\\
We have due to the imposition of $C_I=C_A=0$ 
\ba \label{3.7}
\dot{f} &= & (\lambda^A \{C_A,f\}+\Lambda^I \{C_I,f\})_{C=0,G=0,l=l_0} 
\nonumber\\
&=& \lambda_0^A \{C_A,f\}_{C=0,G=0}+\Lambda_0^I \{C_I,f\}_{C=0,G=0} 
\nonumber\\
&=& (\{\lambda_0^A C_A+\Lambda_0^I C_I,f\})_{C=0,G=0} 
\ea
Now by (\ref{3.4})
\be \label{3.8}
\Lambda_0^I C_I=\delta^I\; y_I+\Lambda_0^I\;[N_I\;^A v_A+h_I]]
\ee
The first term depends only on $z=(x,y)$, the second depends only on $u,x,q,v$
and is linear in $v_A=G_A$. The first term hence has vanishing Poisson brackets 
with both $f$ and $v_A$, the second term has a Poisson bracket with both $f$ and 
$v_A$ which is non vanishing but linear in $G_B$. It thus follows from (\ref{3.4})
and our assumption $\kappa^A=0$
\ba \label{3.9}
\{\lambda_0^A C_A,f\}_{C=0,G=0}
\nonumber\\
&=&
-(\Lambda_0^I\; \{C_I,G_B\}\; (\sigma^{-1})^{BA}\;
\{C_A,f\})_{C=0,G=0}
\nonumber\\
&=&
-(\{\Lambda_0^I\; C_I,G_B\}\; (\sigma^{-1})^{BA}\;
\{C_A,f\})_{C=0,G=0}
\nonumber\\
&=&
-(\{\tilde{h},G_B\}\; (\sigma^{-1})^{BA}\;
\{C_A,f\})_{C=0,G=0}
\ea
where 
\be \label{3.10}
\tilde{h}=\Lambda_0^I h_I
\ee
For the same reason
\be \label{3.11}
\{\Lambda_0^I C_I,f\}_{C=0,G=0}=\{\tilde{h},f\}_{C=0,G=0}
\ee
Hence both terms (\ref{3.9}) and (\ref{3.11}) no longer depend on 
$v,y$ so that $G_A, C_I$ no longer have to be imposed, we only have to 
impose $C_A, G^I$. 
Now we have explictly from (\ref{3.6})
\be \label{3.12}
h=\tilde{h}_{C_A=0, G^I=0}=\tilde{h}(u=g(x,q),x,q,p)_{x=k(\tau)}
\ee
so that 
\be \label{3.13}
\{h,f\}=
\{\tilde{h},f\}_{C_A=0, G^I=0}
-(\frac{\partial \tilde{h}}{\partial u^A}\;
\frac{\partial g^A}{\partial q^{\mathfrak{a}}}\; \frac{\partial f}{\partial p^{\mathfrak{a}}}))_{C_A=0, G^I=0}
=\{\tilde{h},f\}_{C_A=0, G^I=0}
-(\{\tilde{h},v_A\}\;\{g^A,f\})_{C_A=0, G^I=0}
\ee
Since $C_A(u=g(q,x),x,q)\equiv 0$ we have by taking the $q^{\mathfrak{a}}$ 
derivative of this 
identity 
\be \label{3.14}
(\{C_A,f\}-\{C_A,v_B\} \;\{g^B,f\})_{C_A=0, G^I=0}=
(\{C_A,f\}-\sigma_{AB} \;\{g^B,f\})_{C_A=0, G^I=0}=0
\ee
Thus comparing (\ref{3.9}) with the second term in (\ref{3.13}) we arrive at the 
desired result.\\
$\Box$\\

\subsection{The (B,f) description}
\label{s3.2}

In the $(B,f)$ reformulation the $U(1)^3$ model has seven first class constraints:
Three ``Bianchi'' constraints $\hat{C}_j$ linear in $B$ with phase space independent
coefficients as well as the already discussed constraints $\tilde{C}_\mu$ also 
linear in $B$. This makes the discussion even simpler. Using the basis $b^\alpha$ 
from the previous subsection we now subdivide the canonical pairs $(B^a_j, f_a^j)$ 
in just two groups $z=(x^I,y_I)$ and $r=(q^\mathfrak{a},p_\mathfrak{a})$ and write 
the constraints in the form
\be \label{3.15}
C_I=M_I\;^J(x,q)\; y_J+h_I(x,q,p)
\ee
where again the dependence of $M_I\;^J,\;h_I$ on $x,q$ is at most quadratic and 
the dependence of $h_I$ on $p$ is at most linear, however, we will not need this.
Again we assume that the subdivision is such that $M_I\;^J$ is invertible 
on a sufficiently large space of functions as discussed before. We will 
impose gauge fixings of the form $G^I(x)=F^I(x)-\tau^I=$ whose stability under 
$H=\Lambda^I C_I$ leads to the solution for $\Lambda$
\be \label{3.16}
\Lambda_0^J M_J\;^I=(\sigma^{-1})_J\;^I\; \dot{\tau}^J+\kappa^I=:\delta^I,\;
\Lambda_0^I=(M^{-1})_J\;^I\; \delta^J+\hat{\kappa}^I
\ee
where $\sigma_I\;^J=\{y_I,G^J\}$ depends only on $x$, is invertible on a suitable 
space of functions and $\kappa^I$ is in its kernel, thus depending only on $x$.
In particular, $\delta^I$ only depends on $x$. Likewise $\hat{\kappa}$ is in the 
kernel of $M$ and may depend in general on $x,q$. 

Abstracting from the $U(1)^3$ model, we have the general result:
\begin{Theorem} \label{th3.2} ~\\
Let $C_I, G^I$ be as above. Then the physical Hamiltonian reads
\be \label{3.17}
h=(\Lambda_0^I h_I)_{G=0}
\ee
\end{Theorem}
This means that we can completely forget about the constraints with respect to 
the dynamics of the physical degrees of freedom $r$ on the reduced phase space
$C=G=0$. We only need to solve $G=0$ and compute $\Lambda_0$ as in (\ref{3.16})
which depending on the choice of $G$ may be practically conceivable.\\
\\
Proof:\\
Using that 
\be \label{3.18}
\Lambda_0^I C_I=\delta^I y_I+\tilde{h},\; \tilde{h}=\Lambda_0^I h_I
\ee
we have by steps familiar from the previous subsection for $f=f(r)$
\be \label{3.19}
\dot{f}=\{\Lambda_0 C_I,f\}_{C=G=0}=\{\tilde{h},f\}_{G=0}
\ee
Since the solution of $G^I=0$ is of the form $x^I=g^I(\tau)$ and is independent 
of $r$ it follows that we can set $G=0$ also before computing the Poisson bracket.\\
$\Box$

\section{Reduced Phase Dynamics in specific gauges}
\label{s4}

We construct the reduced phase and its physical Hamiltonian in various gauges
which we introduce in the first subsection. In order to keep the technicalities at
a minimum, we restrict attention to linear gauges. In the second subsection, we then apply the theorems of the previous section to compute the corresponding
physical Hamiltonian.

\subsection{Gauge fixing choices}
\label{s4.1}

We split the discussion into the $(A,E)$ and $(B,f)$ descriptions respectively.

\subsubsection{$(A,E)$ description}
\label{s4.1.1}

As the Gauss constraint only involves $E$ we must use a gauge fixing condition
that involves $A$. In the literature of Abelian gauge theories, the Coulomb gauge or axial gauge is popular. Here we consider an extension of both of them to fix also the spacetime diffeomorphism gauge symmetry. In what follows we arbitrarily select the z-coordinate as longitudinal and the x,y coordinates as transversal directions. Transversal spatial indices are now $I,J,\cdots \in \{x,y\}$
while the longitudinal index is denoted by $a=z$.
We consider also ``transversal'' $u(1)^3$ indices $\alpha,\beta, \cdots\in \{1,2\}$ and 
denote the longitudinal one by $3$. The indices $\alpha,\beta,\dots$ like $j,k,\dots$
are moved with the Kronecker $\delta_{\alpha\beta}$, so the index position of $\alpha,\beta,\dots$ is also irrelevant.\\
\\
A.\\
{\it transversally magnetic Coulomb - transversally electric anti-Coulomb - 
longitudinally electric axial gauge (TMC-TEaC-LEA)}\\
The gauge conditions are 
\be \label{4.1}
G^j=\delta^{IJ} \partial_I A_J^j,\;\tilde{G}_j=\epsilon_{IJ} \delta^{IK}
\partial_K E^J_j,\; \tilde{G}_0=E^z_3-f
\ee
where $f$ is a coordinate dependent function possibly also of physical time.
Here $\epsilon_{IJ}=\epsilon^{IJ}$ is the completely skew symbol in two dimensions 
and $\delta^I_J$ is the Kronecker symbol in two dimensions. We will set 
\be \label{4.2}
\partial^I:=\delta^{IJ}\partial_J,\;
\hat{\partial}^I:=\epsilon^{IJ} \partial_J,\;
\hat{\partial}_I:=\epsilon_{IJ} \partial^J
\ee
in what follows to simplify the notation. Clearly, these structures break the 
(spatial) diffeomorphism covariance as they 
should in order to be appropriate gauge fixings in particular of the 
spatial diffeomorphism gauge symmetry.  

We note that the three sets of gauges affect three different sets of canonical 
pairs which have mutually vanishing Poisson brackets among each other, namely 
the transversal Coulomb pair $(\partial^I A_I^j,\Delta^{-1} \partial_I E^I_j)$,
the transversal anti-Coulomb pair $(\hat{\partial}^I A_I^j, 
-\Delta^{-1 }\hat{\partial}_I E^I_j)$ and the longitudinal axial pair
$(A_z^3, E^z_3)$. Here we introduced the two-dimensional Laplacian
\be \label{4.3}
\Delta:=\partial^I \partial_I
\ee
and $\Delta^{-1}$ is a Green function, specifically 
$\Delta^{-1}(x_1,x_2)=(2\pi)^{-1} \ln(||x_1-x_2||)$ where we used the flat 
two dimensional metric.
Note that the transformation to these canonical coordinates is canonical despite the fact that $\Delta$ has a kernel, when 
inverting that transformation the kernel must be taken into account 
using the boundary conditions. 

We note that the Gauss constraint $C_j=\partial_a E^a_j=0$ together with 
the electric gauge conditions imply that the {\it transversal curl} 
$\hat{\partial}_I E^I_j$ vanishes and that the {\it transversal divergence}
$\partial_I E^I_j=-\partial_z E^z_j$ is fixed in terms of $E^z_j$. If we choose
the coordinate function $f$ to be independent of $z$ as we will do then 
in particular $\partial_I E^I_3=\hat{\partial}_I E^I_3=0$. 
Since $\sigma=\mathbb{R}^3$ is simply connected,
$\hat{\partial}_I E^I_3=0$ implies that 
$\delta_{IJ} E^J_3$ is an exact 1-form, that is, $E^I_3=\delta^{IJ}
\partial_J g$ for a certain 0-form $g$ and thus 
$\partial_I E^I_\alpha=\Delta g=0$, hence the $g$ are harmonic 
functions with respect to the $x,y$ dependence. 
By our boundary conditions, $E^I_3$ must decay at infinity and $E^I_3$ is itself
harmonic. 
As is well known, the only smooth
harmonic function on $\mathbb{R}^3$ which decays at infinity is the trivial function. Thus our gauge conditions imply $E^I_3=0$. By the same reasoning from 
$\hat{\partial}_I E^I_\alpha=0$ we infer $E^I_\alpha=\partial_I g_\alpha$ which when
plugged into the Gauss constraint yields $\Delta g_\alpha=-\partial_z E^z_\alpha$.
It follows that $g_\alpha=h_\alpha-\Delta^{-1}\partial_z E^z_\alpha$ where 
$h_\alpha$ is harmonic and thus $E^I_\alpha=\partial^I[h_\alpha-\Delta^{-1} 
\partial_z E^z_\alpha]$. The second term decays at infinity because $E^z_\alpha$ 
does, hence the first term must approach $\delta^I_\alpha$ at infinity. 
It follows that $\partial^I g_\alpha-\delta^I_\alpha$ is a harmonic function vanishing at infinity, hence must vanish itself. Accordingly 
$E^I_\alpha=\delta^I_\alpha-\partial_I \Delta^{-1} \partial_z E^z_\alpha$  
is completely expressed in terms of $E^z_\alpha$. This means 
that the physical degrees of freedom in this gauge correspond to 
$(A_z^\alpha, E^z_\alpha)$. Indeed the condition $\partial^I A_I^j=0$ together
with the boundary conditions implies that $A_I^j$ is a two-dimensional curl
$A_I^j=\hat{\partial}_I g^j$ for certain $g^j$ 
and one would solve the constraints for $g^j$ as well as for $A_z^3$ leaving
$A_z^\alpha$ unconstrained.

It is important to note that the gauge conditions are not in conflict with the 
requirement that the density weight two inverse spatial metric 
$Q^{ab}=E^a_j E^b_k \delta^{jk}$ be non-degenerate.\\
B.\\
{\it Magnetic longitudinal axial - electric tranverse transverse axial 
(MLA-ETTA) gauge}\\
The gauge conditions are
\be \label{4.4}
G^j:=A_z^j,\; G^I_\alpha:=E^I_\alpha-f^I_\alpha
\ee
for certain coordinate dependent functions possibly $f^I_\alpha$ 
also of physical time, e.g. $f^I_\alpha=f \delta^I_\alpha$. 
This gauge is somewhat simpler in that it does not involve derivatives.  
If we assume, as we will, that the $f^I_\alpha$ do not depend on $x,y$ then we have 
from the Gauss constraint $\partial_z E^z_\alpha=0$ which means that 
$E^z_\alpha$ is independent of $z$. Moving to infinity along the $z-$axis at fixed
finite values of $x,y$ this 
can only decay if $E^z_\alpha$ vanishes identically. The third Gauss 
constraint $\partial_z E^z_3+\partial_I E^I_3=0$ is solved 
by $E^z_3=k-\partial_z^{-1} \partial_I E^I_3$ where the Green function 
$\partial_z^{-1}$ is chosen to be 
$(\partial_z^{-1})(z_1,z_2)=\frac{1}{2}{\rm sgn}(z_1-z_2)$ which makes it an 
antisymmetric translation-invariant kernel and $k$ is a function independent of
$z$. Since the second term in $E^z_3$ vanishes at infinity, $k$ must approach 
unity at infinity and $k-1$ must decay. However, since it is independent of $z$ 
this is only possible if in fact $k=1$. It follows that the true degrees of 
freedom in this gauge are $(A_I^3, E^I_3)$ since one would solve the spacetime 
diffeomorphism constraints for $A_I^\alpha$.

Note that this gauge is complementary to the previous one as it leaves 
disjoint sets of canonical pairs as true degrees of freedom.
Again it is easy to verify that the gauge conditions are not in conflict with the requirement that the density weight two inverse spatial metric 
$Q^{ab}=E^a_j E^b_k \delta^{jk}$ be non-degenerate.

\subsubsection{$(B,f)$ description}
\label{s4.1.2}

First of all, we need to know how the canonical variables $(B,f)$ behave in an asymptotically flat spacetime.
By transcribing the boundary conditions imposed on the $(A,E)$ variables, i.e. (\ref{2.1}), to $(B,f)$, we have
\begin{align}
&\epsilon^{abc}\partial_b f_c^j \to F^a_j(n)/r+O(1/r^2),\\
&B_a^j \to \bar{G}_a^j(n)/r^3+O(1/r^4) \label{Asymp B}
\end{align}  
where $\bar{G}_a^j$ are even functions on the asymptotic sphere and $\bar{G}^a_j = \epsilon^{abc}(\delta^d_b - n_b n^d)\frac{\partial G^j_c}{\partial n^d}$. Given (\ref{Asymp B}) and also taking into account the requirement that the symplectic structure must be well defined, the asymptotic behaviour of $f^a_j$ has to be
\begin{equation}\label{Asymp f}
f_a^j \to c^j_a + \bar{F}^j_a(n) + O(1/r) 
\end{equation}
where $c^j_a$ are constants, $\bar{F}^j_a(n)$ are odd functions on the asymptotic $S^2$ and $F^a_j (n)= \epsilon^{abc} \left(\delta^d_b - n_b n^d  \right)\frac{\partial \bar{F}^j_c}{\partial n^d}$. \\
As shown in detail in Appendix \ref{App.A}, there is no well-defined generator for asymptotic symmetries with these boundary conditions. As a result, the use of them is not acceptable. In Appendix \ref{App.A}, an alternative to these boundary conditions is considered, which ultimately leads to the conclusion that Hamiltonian constraint and diffeomorphism constraint are well-defined generators for temporal and spatial translations, respectively. Since here we just deal with the spacetime translations, the lack of generators for boosts and rotations does not have influence on the following calculations. The lack of well-defined generators for boosts and rotations occurs not only in the $(B, f)$ description, but also in that
of $(A, E)$. The latter has been studied in detail in \cite{11}. Note that since the $U(1)^3$ model is not GR,
it is not required to have Poincar\'e group as its asymptotic symmetries.
The appropriate boundary conditions have the fall-off behaviours just the same as (\ref{Asymp B}) and (\ref{Asymp f}) but with different parity conditions
\begin{equation}\label{New PC1}
    \begin{split}
        \bar{F}_a^i\left(-n\right)= \bar{F}_a^i\left(n\right), \; \; \; \bar{G}^a_i\left(-n\right)=-\bar{G}^a_i\left(n\right).
    \end{split}
\end{equation}
Although the chosen boundary conditions do not match those that one would choose
in General Relativity, as the $U(1)^3$ theory is just a toy model for a
generally covariant theory with non-trivial dynamics and an infinite
number of degrees of freedom, we are allowed to exploit the freedom that is allowed in defining the
theory (the choice of boundary condition and choice of polarisation of
the phase space is such an element of freedom) in order to learn as much as possible about the actual theory.
Note that in the case of topologies without boundary, the choice
of boundary conditions is immaterial and the $(A,E)$ description and the $(B,f)$ description are equivalent (see Appendix \ref{App.A}).

As the form of Lagrange multiplier $\Lambda^i$ corresponding to the Bianchi constraint plays an important role in what follows, it is required to find the minimal condition on $\Lambda^i$ ensuring differentiability and convergence of the Bianchi constraint $C_i[\lambda^i]=\int d^3x \;\Lambda^i \partial_a B^a_i$. It is shown in Appendix \ref{App.A} that 
\begin{equation}\label{BC on lambda}
\Lambda^i = \lambda^i r + O(1)
\end{equation}
where $\lambda^i$ are odd functions defined on the asymptotic $S^2$, i.e. 
\begin{equation}\label{PC on lambda}
\lambda^i\left(-n\right)=-\lambda^i\left(n\right)
\end{equation}  
   \\
The gauge fixings discussed below are consistent with the above asymptotic behaviours.
\\
We need also the expression of $H^j_a$ written in terms of $f^j_a$ which can be easily derived from combining (\ref{2.7}) and  $E^a_j=\delta^a_j +\epsilon^{abc} \partial_b f_c^j$,
\begin{align}
H^j_a = \delta^j_a + \epsilon^{jkl}(\partial_a f^l_b - \partial_b f^l_a)[\delta^b_k + \frac{1}{2}\epsilon^{bde}(\partial_d f^k_e)]
\end{align}
\\
As in the $(A,E)$ description, we consider z-coordinate as the longitudinal direction and $I, J, \dots \in \{x, y\}$ as the transversal ones. Similarly, in the internal indices, 3 is considered as the longitudinal and $\alpha, \beta, \dots \in \{1, 2\}$ as the transversal directions.\\
In what follows, appearing several differential operators and functions in a row means that each operator acts on all the operators and functions existing in its right side, for instance if we have an operator as $X:= \partial_x g \partial_y^{-1} f + h$ in which $f, g, h$ are functions, then the action of $X$ on a function $u$ would be $X u= \partial_x (g \partial_y^{-1} (f u)) + h u $.
\\
\\
A.\\
{\it Electric transverse axial - longitudinally electric axial (ETA-LEA) gauge}\\
The gauge conditions are 
\begin{align}\label{ETA-LEA}
& G_I^j =\; f^j_I - \sigma^j_I, \; \; \; G =\; f^3_z - \sigma
\end{align}
in which $\sigma^j_I$ and $\sigma$ are phase space independent functions which may depend only on the physical time, $\tau$. If we assume, as we will, that $\sigma^j_I$ and $\sigma$ do not depend on the spatial coordinates, then not only will the calculations be simpler but also we can make sure that (\ref{ETA-LEA}) is compatible with (\ref{Asymp f}) and (\ref{New PC1}). Applying the gauge fixing (\ref{ETA-LEA}) on $H^j_a$ results in
\begin{align}
H^j_x =& \delta^j_x + \epsilon^{jkl}(\partial_x f^l_b - \partial_b f^l_x)[\delta^b_k + \frac{1}{2}\epsilon^{bde}(\partial_d f^k_e)]
=\delta^j_x + \epsilon^{j3l}(\partial_x f^l_z)
\nonumber \\
H^j_y =& \delta^j_y + \epsilon^{jkl}(\partial_y f^l_b - \partial_b f^l_y)[\delta^b_k + \frac{1}{2}\epsilon^{bde}(\partial_d f^k_e)]
=
\delta^j_y + \epsilon^{j3l}(\partial_y f^l_z)\nonumber \\
H^j_z =& \delta^j_z + \epsilon^{jkl}(\partial_z f^l_b - \partial_b f^l_z)[\delta^b_k + \frac{1}{2}\epsilon^{bde}(\partial_d f^k_e)]
=
\delta^j_z - \epsilon^{j1l}(\partial_x f^l_z) - \epsilon^{j2l}(\partial_y f^l_z) + \epsilon^{jkl}(\partial_y f^l_z)(\partial_x f^k_z)
\end{align}
Thus, $H_I^3= H^\alpha_z = 0$ and the non-vanishing ones are
\begin{align}\label{Hs in ETA-LEA}
H^1_x =& \; 1- \partial_x f^2_z, \; \; \; H^2_x = \partial_x f^1_z,\nonumber \\
H^1_y =& \; - \partial_y f^2_z, \; \; \; \; \; H^2_y =1 + \partial_y f^1_z, \nonumber \\
H^3_z =& \; 1 - \partial_x f^2_z + \partial_y f^1_z - (\partial_x f^2_z)(\partial_y f^1_z) + (\partial_x f^1_z)(\partial_y f^2_z)\nonumber \\
=& \; H^1_x H^2_y - H^2_x H^1_y 
\end{align}
With the $H^a_j$ evaluated at the gauge cut, the constraints are much simpler to solve for $B^I_i$ and $B^z_3$. In fact, the matrix representation of (\ref{3.15}) in this gauge is
\begin{equation}\label{B MTA-MLA}
\left[ {\begin{array}{ccccccc}
   \partial_x & 0 & 0 & \partial_y & 0 & 0 & 0 \\
   0 & \partial_x & 0 & 0 & \partial_y & 0 & 0 \\
   0 & 0 & \partial_x & 0 & 0 & \partial_y & \partial_z \\
   0 & 0 & -H^2_x & 0 & 0 & -H^2_y & 0 \\
   0 & 0 & H^1_x & 0 & 0 & H^1_y & 0 \\
   H^2_x & -H^1_x & 0 & H^2_y & -H^1_y & 0 & 0 \\
   H^1_x & H^2_x & 0 & H^1_y & H^2_y & 0 & H^3_z \\
  \end{array} } \right]
  \left[ {\begin{array}{c}
   B^x_1  \\
   B^x_2  \\
   B^x_3  \\
   B^y_1  \\
   B^y_2  \\
   B^y_3  \\
   B^z_3  \\
  \end{array} } \right]
  +
  \left[ {\begin{array}{c}
   \partial_z B^z_1  \\
   \partial_z B^z_2  \\
   0  \\
   B^z_2 H^3_z  \\
   -B^z_1 H^3_z  \\
   0  \\
   0  \\
  \end{array} } \right]=0
\end{equation} 
\\
Note that the existence of the solutions for this system of equations is essentially dependent on the non-vanishing condition of $H^3_z$ which is guaranteed in the MTA-MLA gauge (see Appendix \ref{App. B}). According to (\ref{2.7}) and the fact that in this gauge $H^1_z=H^2_z=0$, vanishing of $H^3_z$ would lead to a degenerate $E$, which contradicts the requirement that the spatial metric must be non-degenerate. Thus, $H^3_z\neq 0$ everywhere.
\\
As one can see in detail in Appendix \ref{App. B}, solving the system (\ref{B MTA-MLA}) using the boundary conditions completely gives us $(B^I_i, B^z_3)$ in terms of $B^z_\alpha$.
Therefore, the true degrees of freedom in this gauge are $(B^z_\alpha, f^\alpha_z)$.
\\

In this gauge, one can easily see that
\begin{equation}
E^a_i=
\left[ {\begin{array}{ccc}
   H^2_y & -H^1_y & 0 \\
   -H^2_x & H^1_x & 0  \\
   0 & 0 & 1  \\
  \end{array} } \right]
\end{equation}
Therefore the gauge conditions are not in conflict with the requirement that the density weight two inverse spatial metric $Q^{ab}=E^a_j E^b_k \delta^{jk}$ be non-degenerate.
\\
\\
B.\\
{\it Electric transverse transverse axial- electric longitudinal axial (ETTA-ELA) gauge}
\\
The gauge conditions are
\begin{equation}\label{the second gauge conditions}
G^\alpha_I= f^\alpha_I -\sigma^\alpha_I, \;\;\;\; G^j= f^j_z -\sigma^j 
\end{equation}
where $\sigma^\alpha_I$ and $\sigma^j$ are phase space independent functions which may depend only on the physical time, $\tau$. In order to make the following calculations simpler we assume that $\sigma^\alpha_I$ and $\sigma^j$ do not depend on the spatial coordinates. It is assured that (\ref{ETA-LEA}) is totally compatible with (\ref{Asymp f}) and (\ref{New PC1}).
A straightforward calculation shows that in this gauge $H^j_a$ has a very simple form, that is
\begin{align}\label{Hs in the second gauge fixing}
&H^1_x= H^2_y = 1+ \partial_x f^3_y - \partial_y f^3_x\nonumber\\
&H^1_z = \partial_z f^3_y - \partial_y f^3_z=\partial_z f^3_y  \nonumber\\
&H^2_z = \partial_x f^3_z - \partial_z f^3_x =  - \partial_z f^3_x \nonumber\\
&H^2_x = H^3_x = H^1_y = H^3_y =0, \;\;\;\; H^3_z=1
\end{align} 
With these $H$'s, we can see that the system of equations that has to be solved here is even simpler than that of the previous gauge, i.e. (\ref{B MTA-MLA}). In fact, the matrix representation of (\ref{3.15}) in this gauge is
\begin{equation}\label{B ETTA-ELA}
\left[ {\begin{array}{ccccccc}
   \partial_x & 0 & \partial_y & 0 & \partial_z & 0 & 0 \\
   0 & \partial_x & 0 & \partial_y & 0 & \partial_z & 0 \\
   0 & 0 & 0 & 0 & 0 & 0 & \partial_z \\
   0 & 0 & 0 & 0 & 0 & 1 & -H^2_z \\
   0 & 0 & 0 & 0 & -1 & 0 & H^1_z \\
   0 & -H^1_x & H^1_x & 0 & H^2_z & -H^1_z & 0 \\
   H^1_x & 0 & 0 & H^1_x & H^1_z & H^2_z & 1 \\
  \end{array} } \right]
  \left[ {\begin{array}{c}
   B^x_1  \\
   B^x_2  \\
   B^y_1  \\
   B^y_2  \\
   B^z_1  \\
   B^z_2  \\
   B^z_3  \\
  \end{array} } \right]
  +
  \left[ {\begin{array}{c}
   0  \\
   0  \\
   \partial_x B^x_3 +\partial_y B^y_3  \\
   -H^1_x B^y_3  \\
   H^1_x B^x_3  \\
   0  \\
   0  \\
  \end{array} } \right]=0
\end{equation} 
\\
Note that the existence of the solutions for (\ref{B ETTA-ELA}) basically depends on the non-vanishing condition of $H^1_x$ which is guaranteed in the ETTA-ELA gauge (see Appendix \ref{App. B}). According to (\ref{2.7}) and the fact that in this gauge $H^3_I=H^2_x=H^1_y=0$, vanishing of $H^1_x$ would lead to a degenerate $E$, which contradicts the requirement that the spatial metric must be non-degenerate. Thus, $H^1_x\neq 0$ everywhere.
\\
In Appendix \ref{App. B}, It is shown that by solving the system (\ref{B ETTA-ELA}) using the boundary conditions, one can express $(B^I_\alpha, B^z_j)$ completely in terms of $B^I_3$.
Therefore, the true degrees of freedom in this gauge are $(B^I_3, f^3_I)$.
\\

In this gauge, one can easily see that
\begin{equation}
E^a_i=
\left[ {\begin{array}{ccc}
   1 & 0 & -H^1_z \\
   0 & 1 & -H^2_z  \\
   0 & 0 & H^1_x  \\
  \end{array} } \right]
\end{equation}
Therefore, again, the gauge conditions are not in conflict with the requirement that the density weight two inverse spatial metric $Q^{ab}=E^a_j E^b_k \delta^{jk}$ be non-degenerate.
\subsection{Reduced phase space dynamics in various gauges}
\label{s4.2}

We will now discuss the reduced phase space dynamics in various gauges 
and using 
different pairs of canonical coordinates. 

\subsubsection{MLA-ETTA gauge in $(A,E)$ description}
\label{s4.2.1}
We verify that the assumptions of theorem \ref{th3.1} apply:\\
First recall that in this gauge condition we have $w=(E^z_j, A_z^j)$, $z=(E^I_\alpha, A_I^\alpha)$ and $r=(E^z_\alpha, A_z^\alpha)$.
We interpret (minus) the $A_z^j$ as the momenta $v_A$ and   
(minus) the $A_I^\alpha$ as the momenta $y_I$. Likewise the Gauss constraints 
are considered as the $C_A$ and the spacetime constraints as the $C_I$.  

We have 
\be \label{4.5}
\{C(\lambda), G_j\}=
\{C(\lambda), A_z^j\}=-\kappa \partial_z \lambda^j 
\ee
which has as kernel the space of functions that are independent of $z$. If we 
consider as space of functions $\lambda^j$ those that decay at 
infinity at least as $1/r^2$ this means that the kernel vanishes and 
the matrix $\sigma_{AB}$ considered as the integral kernel $\{C_j(x),G_j(y)\}$ 
is regular. 

Next we have 
\be \label{4.6}
\tilde{C}(U,F)
=\int\; d^3x\; B^a_j\; (F H_a^j+\epsilon^{jkl} H_a^k U^l)
=: \int\; d^3x\; B^a_j\; u_a^j
=: \int\; d^3x\; A_a^j v^a_j
\ee
with 
\be \label{4.7}
v^a_j=\epsilon^{abc} \partial_b u_c^j,\; u_a^j=F H_a^j+\epsilon^{jkl} H_a^k U^l
\ee
Geometrically, the relation between the density weight minus smearing functions 
$F, U^j$ one and the 
density weight zero lapse and shift functions 
$N, N^a$
is that
\be \label{4.8}
F=N |\det(E)|^{-1/2},\;\;
U^j E^a_j=N^a
\ee
i.e. $N^a$ is an ``electric shift'' \cite{7} if $U,F$ are considered as phase space 
independent\footnote{The lapse function $N$ and shift vector $N^a$ can be considered as Lagrange multipliers of the Hamiltonian and diffeomorphism constraints, respectively. Alternatively, one can consider the density weight $-1$ versions of them as the smearing functions of the constraints, i.e., $N |\det(E)|^{-1/2}$ and $N^a E_a^j$, respectively. This is the geometric origin of (\ref{4.8}).}.

Obviously 
\be \label{4.9} 
\{\tilde{C}(U,F),E^a_j\}=-\kappa v^a_j
\ee
and we must verify that, given asymptotically flat 
boundary conditions, $-\kappa v^\alpha_I=\dot{\tau}^\alpha_I$ implies unique values 
for $F,U$ in order that the gauge fixing be admissible. It will be sufficient to 
verify this at the gauge cut $C_j=\partial_a E^j=0,\; E^I_\alpha=\tau^I_\alpha$ 
because then by continuity, it will hold in a neighbourhood of the gauge cut by the usual reasoning of the implicit function theorem (to make this precise, the phase space 
should be given a suitable (Banach) manifold topology).    

To be specific we consider $\tau^I_\alpha=f \delta^I_\alpha$ where $f$ is 
at most depending on the physical time. In fact this is compatible with asymptotic 
flatness if we further specify $f\equiv 1$, but let us be more general for the moment.
As already mentioned, from the Gauss constraint we then obtain 
$\partial_z E^z_\alpha=0$ which implies $E^z_\alpha=0$ if we use the boundary 
conditions. Furthermore $\partial_a E^a_3=0$ is uniquely solved by 
$E^z_3=1-\partial_z^{-1} \partial_I E^I_3$ so that $E^z_3$ is completely
determined by $E^I_3$. We evaluate the $H_a^j$ at the gauge cut
\ba \label{4.10}
H_z^3 &=&\det(\{E^I_\alpha\})=f^2
\nonumber\\
H_I^3 &=& \epsilon^{\alpha\beta} \epsilon_{IJ} E^I_\beta E^z_\beta=0
\nonumber\\
H_z^\alpha &=& \epsilon_{IJ} \epsilon^{\alpha\beta} E^I_\beta E^J_3
=-f \delta_I^\alpha\; E^I_3
\nonumber\\
H_I^\alpha &=& \epsilon_{IJ} \epsilon^{\alpha\beta}[E^J_\beta E^z_3-E^J_3 E^z_\beta]
=f \delta_I^\alpha E^z_3
\ea
and find with 
$\hat{U}_\alpha=\epsilon_{\alpha\beta} U^\beta,\;
\hat{U}^\alpha=\epsilon^{\alpha\beta} U_\beta$
\ba \label{4.11}
u_z^3 &=& F H_z^3+\epsilon_{\alpha\beta} H_z^\alpha U^\beta
=F H_z^3 + H_z^\alpha \hat{U}_\alpha 
\nonumber\\
u_z^\alpha &=& F H_z^\alpha+ \epsilon_{\alpha\beta}(H_z^\beta U^3-H_z^3 U^\beta)
=[F\delta^\alpha_\beta+U^3 \epsilon_{\alpha\beta}]H_z^\beta-H_z^3 \hat{U}^\alpha
\nonumber\\
&=:& \Sigma^\alpha_\beta H_z^\beta-H_z^3 \hat{U}^\alpha
\nonumber\\
u_I^3 &=&F H_I^3+\epsilon^{\alpha\beta} H_I^\alpha U^\beta
=H_I^\alpha \hat{U}_\alpha
\nonumber\\
u_I^\alpha &=& F H_I^\alpha+\epsilon^{\alpha\beta} (H_I^\beta U^3-H_I^3 U^\beta)
=\Sigma^\alpha_\beta H_I^\beta
\ea

The stability condition resulting from (\ref{4.9}) is 
\be \label{4.12}
-\dot{f}/\kappa \delta^I_\alpha
=v^I_\alpha=\epsilon^{IJ}[\partial_J u_z^\alpha-\partial_z u_J^\alpha]
\ee
This can be transformed into
\be \label{4.13}
\partial_I u_z^\alpha=\frac{\dot{f}}{\kappa} \epsilon_{IJ}\delta^J_\alpha
+\partial_z u_I^\alpha
\ee
Let us introduce 
\be \label{4.14}
a:=f E^z_3 F,\; b:= f E^z_3 U^3
\ee
then using (\ref{4.10}), (\ref{4.11}) 
\be \label{4.15}
u_I^\alpha=a\delta^\alpha_I+ b\epsilon_{\alpha\beta} \delta^\beta_I
\ee
and the four equations (\ref{4.13}) can be disentangled 
\be \label{4.16}
\partial_x u_z^1=\partial_y u_z^2=\partial_z a;\;\;
\partial_y u_z^1=-\partial_x u_z^2=\partial_z b-\frac{\dot{f}}{\kappa}
\ee
We thus find that $(u_z^1, u_z^2)$ forms a Cauchy-Riemann pair, i.e. 
$w=u_z^1+iu_z^2$ is a holomorphic function in $x+iy$. On the other hand
from (\ref{4.10}) and (\ref{4.11}) we infer that $u_z^\alpha$ is a bounded function 
if $U^j, F, E^I_3$ are or that it decays at least as $r^{-1}$ if we allow 
$N, N^a$ and thus $F, U^j$ (recall (\ref{4.8}))
to approach asymptotically at most a constant and zero
respectively. In the former case, we infer $u_z^\alpha=$const. by Liouville's 
theorem. In the latter case we obtain immediately $u_z^\alpha=0$ because 
a holomorphic function decays at most in $x^2+y^2$ but not in $r^2$. 
In either case 
\be \label{4.17}
\partial_z a=0=\partial_z b-\frac{\dot{f}}{\kappa}
\ee
which is solved by 
\be \label{4.18}
a=a_0(x,y),\; b=b_0(x,y)+\frac{\dot{f}}{\kappa} z
\ee
Comparing with (\ref{4.14}) we see that a approaches asymptotically a constant
up to $O(1/r)$ corrections and that $b$  decays at least as $1/r$. Since a function 
of $x,y$ cannot decay in $z$ as $r$ does, we must have in fact $a_0=$const. and 
$b_0=\dot{f}=0$. Furthermore by comparing the asymptotic values we find $a_0=f$
and we conclude $f=1$ from $E^a_j=\delta^a_j+O(1/r)$. Accordingly
\be \label{4.19}
F=\frac{1}{E^z_3},\; U^3=0
\ee
and thus 
\be \label{4.20}
u_z^\alpha=F H_z^\alpha -H_z^3 \hat{U}^\alpha=c^\alpha
\ee
where $c^\alpha=$const. However, since $H_z^3=1$ a non vanishing constant is 
incompatible with a decaying $U^\alpha$ so that in fact $u_z^\alpha=0$ and 
\be \label{4.21}
\hat{U}^\alpha=\frac{H_z^\alpha}{E^z_3}=-\delta^\alpha_I\frac{E^I_3}{E^z_3}
\ee
We conclude that the gauge fixing has resulted in a unique solution for 
$F, U^j$ and thus was admissible.\\
\\
We now apply theorem \ref{th3.1} to compute the physical Hamiltonian $h$ which
depends only on the physical degrees of freedom. We thus decompose 
\be \label{4.22}
\tilde{C}(F,U)=\int\; d^3x\; [v^I_\alpha A_I^\alpha+v^z_j A_z^j+v^I_3 A_I^3] 
\ee
and identify the first term with $\Lambda^I M_I\;^J y_I$, the second term 
with $\Lambda^I N_I\;^A v_A$ and the third term with $\Lambda^I h_I$ where 
$\Lambda^I$ corresponds to $(F,U)$. Upon integrations by parts we have 
\ba \label{4.23}
\Lambda^I h_I
&=& \int\;d^3x\; \epsilon^{IJ}(\partial_J u_z^3-\partial_z u_J^3) A_I^3
\nonumber\\
&=& \int\;d^3x\; 
[u_z^3 (\epsilon^{IJ} \partial_I A_J^3)+u_I^3 (-\epsilon^{IJ} \partial_z A_J^3)]
\nonumber\\
&=&\int\;d^3x\; 
[u_z^3 B^z_3+u_I^3 B^I_3]
\nonumber\\
&=&\int\;d^3x\; 
[F (H_z^3 B^z_3+ H_I^3 B^I_3)+\hat{U}^\alpha (H_z^\alpha B^z_3+H_I^\alpha B^I_3)]
\nonumber\\
&=&\int\;d^3x\; 
[F H_a^3 B^a_3+\hat{U}_\alpha H_a^\alpha B^a_3]
\ea
where $B^I_3$ is to be evaluated in the magnetic longitudinal axial gauge 
$A_z^3=0$. Note that in fact (\ref{4.23}) does not depend on $U^3$ even before using the constraints, the gauge conditions, and the gauge fixed Lagrange 
multipliers.

According to theorem \ref{th3.1} we only need to impose the gauge 
$E^I_\alpha=\delta^I_\alpha$ and use the solution $E^z_\alpha=0, \; 
E^z_3=1-\partial_z^{-1} \partial_I E^I_3$ of the Gauss constraint as well as the 
solution for $\Lambda_0^I=(F_0,U_0)$ at these values which are displayed in 
(\ref{4.19}), (\ref{4.21}) and then insert those into (\ref{4.23}). This yields
the final expression
\ba \label{4.24}
h&=& \int\; d^3x\; [B^z_3(F+H_z^\alpha \hat{U}_\alpha)+B^I_3 H_I^\alpha 
\hat{U}_\alpha]
\nonumber\\
&=& \int\; d^3x\; [B^z_3(\frac{1}{E^z_3}+\delta_{IJ}\frac{E^I_3 E^J_3}{E^z_3})
-B^I_3 \delta_{IJ} E^J_3]
\nonumber\\
&=&
\int\; d^3x\; A_I^3\; \epsilon^{IJ} 
[\partial_J(\frac{1+\delta_{IJ} E^I_3 E^J_3}{E^z_3})-\partial_z \delta_{JK} E^K_3]
\ea
As expected, the physical Hamiltonian (\ref{4.24}) has the following features:\\
I. Linearity in momentum $A_I^3$.\\
II. Non-polynomiality in the configuration variable $E^I_3$.\\
III. Spatial non-locality.\\
The latter two properties are due to the appearance of $1/E^z_3$ with 
$E^z_3=1-\partial_z^{-1} \partial_I E^I_3$. Note that the term 
$1+\delta_{IJ} E^I_e E^J_3$ is meaningful from a dimensional point of view 
because $E^a_j$ is dimension-free and $A_a^j$ has dimension cm$^{-1}$. Thus the 
Hamiltonian density has dimension $cm^{-2}$ and to turn it into a quantity with dimension of energy we have to divide $h$ by $\kappa$ (if time is multiplied by the speed of light). In fact, we should have worked all the time with constraints rescaled by $1/\kappa$ as they would naturally appear out of an action \cite{15}. The fact that the three individual
terms in the Hamiltonian have different density weights is due to the gauge fixing condition $E^I_\alpha=\delta^I_\alpha$ which breaks the density weight of $E^I_\alpha$.\\
\\
We can now derive the physical equations of motion and indicate strategies for 
how to solve them, despite the complexity of the Hamiltonian (\ref{4.24}).
Beginning with $E^I_3$ we have, using $\{E^I_3(x),A_J^3(y)\}=\kappa \delta^I_J 
\delta(x,y)$ 
\be \label{4.25}
\dot{E}^I_3=\{\frac{h}{\kappa},E^I_3\}
=-\epsilon^{IJ} 
[\partial_J(\frac{1+\delta_{IJ} E^I_3 E^J_3}{E^z_3})-\partial_z \delta_{JK} E^K_3]
\ee
As expected, due to linearity of $h$ in $A$, the equation of motion for 
$E^I_3$ is of first order in time and closes on itself, i.e its time derivative 
no longer involves $A_I^3$ and one can study (\ref{4.25}) completely independently
of the equation of motion for $A_I^3$. Next we have 
\be \label{4.26}
\dot{A}_I^3(x)=\{\frac{h}{\kappa},A_I^3(x)\}
=(-B^I_3+2 B^z_3\frac{E^I_3}{E_z^3})(x)
-\int\; d^3y\; (B^z_3 \frac{1+\delta_{JK} E^J_3 E^K_3}{(E^z_3)^2})(y)\;
\frac{\delta E^z_3(y)}{\delta E^I_3(x)}
\ee
We have for any smearing function $s$ using 
$E^z_3=-\partial_z^{-1} \partial_J E^J_3$, integrations by parts and 
the antisymmetry of the integral kernel of $\partial_z^{-1}$ as well as 
$[\partial_z^{-1},\partial_I]=0$ 
\be \label{4.27}
\int\; d^3y \;s(y) \frac{\delta E^z_3(y)}{\delta E^I_3(x)}
=\int\; d^3y \;(\partial_z^{-1} \partial_J s)(y) 
\frac{\delta E^J_3(y)}{\delta E^I_3(x)}
=(\partial_z^{-1} \partial_J s)(x)  
\ee
whence 
\be \label{4.28}
\dot{A}_I^3(x)=\{\frac{h}{\kappa},A_I^3(x)\}
=-B^I_3+2 B^z_3\frac{E^I_3}{E_z^3}
-\partial_z^{-1} \partial_I (B^z_3 \frac{1+\delta_{JK} E^J_3 E^K_3}{(E^z_3)^2})
\ee
Note the abbreviations $B^I_3:=-\epsilon^{IJ}\partial_z A_J^3,\;
B^z_3:= \epsilon^{IJ} \partial_I A_J^3$.

Equations (\ref{4.25}), (\ref{4.28}) suggest the following solution strategy:
First solve (\ref{4.25}) which is independent of $A$. Then plug that solution 
into (\ref{4.28}) which is then a linear integro-differential equation system of 
first order in all derivatives and the anti-derivative $\partial_z^{-1}$. 
Turning to the first task we introduce the {\it divergence} and {\it curl}
of $E^I_3$
\be \label{4.29}
D:=\partial_I E^I_3,\;C:=\epsilon^{IJ} \partial_I (\delta_{JK} E^K_3) 
\ee
and decompose 
\be \label{4.30}
E^I_3=\Delta^{-1}[\delta^{IJ} \partial_J D-\epsilon^{IJ} \partial_J C]
\ee
where $\Delta$ is the transversal Laplacian and we used the boundary conditions 
($E^I_3$ must vanish as $1/r$) to exclude a non-vanishing kernel of $\Delta$.
Taking the divergence and curl of (\ref{4.25}) we find 
\be \label{4.30}
-\dot{D}=-\partial_z C,\;\;
-\dot{C}=-\Delta[\frac{1+\delta_{IJ} E^I_3 E^J_3}{E^z_3}]+\partial_z D 
\ee
Note that if we would drop the non-linear interaction term in the second equation 
of (\ref{4.30}) and would iterate it we would find 
\be \label{4.31}
\ddot{D}+\partial_z^2 D=\ddot{C}+\partial_z^2 C=0
\ee
which is a ``Euclidean'' wave operator restricted to the $z$ direction. We interpret
this to express the closeness of the model to Euclidean gravity. Continuing
with (\ref{4.30}) we remember that $D=-\partial_z E^z_3$ so that the first
equation in (\ref{4.30}) implies (again possible integration constants must 
vanish)  
\be \label{4.32}
C=-\dot{E}^z_3
\ee
This means that all of $E^I_3$ can be written just in terms of $F:=E^z_3$
\be \label{4.33}
E^I_3=-\Delta^{-1}[\delta^{IJ} \partial_J \partial_z-\epsilon^{IJ} \partial_J 
\partial_t]F
\ee
which allows to write the second equation in (\ref{4.30}) just in terms of 
$F$
\be \label{4.34}
\ddot{F}+\partial_z^2 F=
-\Delta[\frac{1+\delta_{IJ} E^I_3 E^J_3}{F}]
\ee
where (\ref{4.33}) is to be used on the r.h.s.. Equation (\ref{4.34}) 
has the undesirable feature that it is non-polynomial (due to $1/F$) and spatially 
non-local (due to $\Delta^{-1}$ in (\ref{4.33})). We can get rid of the first 
feature by multiplying (\ref{4.34}) with $F^3$ because the Laplacian on the r.h.s.
produces at most factors of $1/F^3$. We can get rid off the second feature by 
introducing $G=\Delta^{-1} F,\; F=\Delta G$. Accordingly, we write
\be \label{4.35}
E^I_3=-[\delta^{IJ} \partial_J \partial_z-\epsilon^{IJ} \partial_J 
\partial_t]G
\ee
and 
\be \label{4.36}
(\Delta G)^3[\partial_t^2+\partial_z^2](\Delta G)
=-(\Delta G)^3[\Delta[\frac{1+\delta_{IJ} E^I_3 E^J_3}{\Delta G}]
\ee
where (\ref{4.35}) has to be substituted. Thus (\ref{4.36}) is a polynomial
in $G$ of degree four and involves time derivatives up to second order, spatial 
derivatives up to order four. To see this we write with 
$K:=\delta_{IK} E^I_3 E^J_3$ 
\ba \label{4.37}
F^3\Delta(\frac{1+K}{F})
&=&
F^3(\frac{\Delta K}{F}-2\frac{\partial_I K}{F^2} (\partial^I F)+
(1+K)[-\frac{\Delta F}{F^2}+2\frac{(\partial_I F)(\partial^I F)}{F^3}])
\nonumber\\
&=&
F^2(\Delta K)-2 F (\partial_I K) (\partial^I F)+
(1+K)[-F (\Delta F)+2 (\partial_I F)(\partial^I F)]
\ea
We leave the further analysis of (\ref{4.36}) for future work. It is remarkable that half of the equations of motion can be encoded just in terms of a {\it single} PDE! The fact that the theory is self-interacting is expressed by the fact that this PDE is far from linear and of spatial degree higher than two (namely four) but still of temporal degree at most two. In particular,
it is a quasi-linear equation with respect to the highest time derivatives.\\
\\
As far as Hilbert space representations are concerned that support (some ordering 
of) $h$ as a densely defined operator as well as the corresponding spectral problem
consider the possibility of a representation in which $E^I_3$ acts as a 
multiplication operator, i.e. ${\cal H}=L_2(\widehat{{\cal E}},d\mu)$ where 
$\widehat{{\cal E}}$ is a 
suitable distributional extension of the set $\cal E$ of the classical fields 
$E^I_3$ and $\mu$ a probability measure 
thereon. Then $A_I^3=i\ell_P^2 \delta/\delta E^I_3+D_I(E)$ where 
$D_I$ is chosen as to make $A_I^3$ a symmetric operator valued distribution and 
$\ell_P^2=\hbar \kappa$ is the Planck area.
Then for any symmetric ordering of $h$, after reordering in such a way that the
functional derivatives are acting directly on the Hilbert space vector $\psi\in
{\cal H}$ we find that $h$ acts as
as
\be \label{4.38}
(h\psi)[E]=i\ell_P^2\int\; d^3x\; V^I_3(E(x))\; \frac{\delta \psi}{\delta E^I_3(x)}
+U[E]\psi[E]
\ee
where the potential term $U[E]$ acts as a multiplication operator and stems 
from reordering $h$ into the form displayed as well as from the contribution 
$V^I D_I$ where 
\be \label{4.39}
V^I_3(E(x))=\epsilon^{IJ}[\partial_J(\frac{1+K}{F})-\partial_z E^J_3]
\ee
The reordering of $h$ in the form displayed produces in general singularities 
in the form of (derivatives of) $\delta$ distributions evaluated at zero 
which need to be regularised and which guide the choice of $\mu$ and thus 
$D_I$ to cancel them. Also the form of $V^I, U$ may raise non-trivial domain questions.

We can now recast the spectral problem for $h$ into the form 
\be \label{4.40}
i\ell_P^2 <V^I,\frac{\delta}{\delta E^I_3}> \; \psi=(\lambda-U)\;\psi
\ee
where $<.,.>$ denotes the inner product on $L_2(d^3x, \mathbb{R}^3)$ 
or equivalently with the WKB Ansatz $\psi=\exp(-i \frac{S}{\ell_P^2})$
\be \label{4.40}
<V^I,\frac{\delta S}{\delta E^I_3}>=(\lambda-U)
\ee
Equation (\ref{4.40}) has the form of a {\it linear functional partial differential
equation of first order}, that is, the infinite-dimensional analogue of the well-known case of a linear partial differential equation of first order. 
This is the simplest type of first-order FPDE (functional PDE) that one can imagine,
it is not even quasi-linear (which would allow $V^I, U$ to depend on $S$) or 
even non-linear (which would allow $V^I,U$ to depend on $S,\delta S/\delta E^I_3$).
Note also that in contrast to Hamiltonians with at least quadratic dependence on
on the momenta, the WKB Ansatz leads to a Hamilton-Jacobi equation
(\ref{4.40}) which in this case is {\it exact}. 

One can then try to solve (\ref{4.40}) by a functional version of the 
method of characteristics \cite{16}
which brings out the closeness of the spectral problem (\ref{4.40}) 
to the solution of the classical equations of motion (\ref{4.25}) in this case.
To that end, we solve the equations of motion (\ref{4.25}), i.e. we determine 
the integral curves of the vector field $V^I$. Suppose that we found the maximal
unique  
solution $X^I(t,x;G)$ given prescribed initial data $X^I(0,x)=G^I(x)$ where 
$G$ ranges in some submanifold $\Sigma\subset {\cal E}$ 
of co-dimension one which is everywhere transversal to the vector field $V$. 
We also solve 
\be \label{4.41}
\dot{s}(t)=\lambda-U[E=X(t,.;G)]  
\ee
with initial condition $s(0)=s_0[G]$ leading to a unique maximal solution
$s(t;G)$. The transversality of $\Sigma$ to the flow lines of $V$ implies that 
at least for small $t$ i.e. close to $\Sigma$ we may invert the equation
$E^I_3(x)=X^I(t,x;G)$ for 
\be \label{4.42}
t=\tau[E],\;G^I(x)=\sigma^I(x;E)
\ee
and then 
\be \label{4.43}
S[E]:=s(t,G)_{t=\tau, G=\sigma}
\ee
solves (\ref{4.40}) with boundary condition $S[G]=s_0[G]$ i.e. 
$S_{|\Sigma}=s_0$. 
 
While the precise technical implementation of these steps may be quite involved,
they are, remarkably, of much lower complexity than one might have feared.
We leave the details of this programme for future work.

\subsubsection{TMC-TEaC-LEA gauge in $(A,E)$ description}
\label{s4.2.2}

We verify that the assumptions of theorem \ref{th3.1} apply:\\
We interpret (minus) the $v^j:=\partial^I A_I^j$ as the momenta $v_A$ and   
(minus) the $y^j:=\hat{\partial}^I A_I^j, A^3_z$ as the momenta $y_I$. The configuration variables corresponding to these momenta are $u_j:= \Delta^{-1}\partial_I E^I_j$, $x_j:= \Delta^{-1}\hat{\partial}_I E^I_j$ and $E^z_3$, respectively. Likewise the Gauss constraints are considered as the $C_A$ and the spacetime constraints as the $C_I$. In what follows, we work with the canonical variables $w=(u_j, v^j), z=\{(x_j, y^j), (E^z_3, A^3_z)\}$ and $r=(E^z_\alpha, A^\alpha_z)$ where the latter plays the role of our degrees of freedom. Thus, it is required to express $E^I_j$ and $A^j_I$ in terms of $u_j, x_j$ and $v^j, y^j$, respectively
\begin{align}
E^I_j &= \delta^I_j + \partial^I u_j + \hat{\partial}^I x_j\\
A^j_I &= \Delta^{-1} \left( \partial_I v^j + \hat{\partial}_I y^j \right)
\end{align}
And the constraints in terms of the canonical variables are
\begin{align}\label{Cj}
\tilde{C}_j &= \epsilon_{jkl}(B^I_k H^l_I + B^z_k H^l_z)\nonumber\\
&= 
\epsilon_{jkl}\left(\epsilon^{IJ} H^l_I (\partial_J A^k_z - \partial_z (\partial_J v^k + \hat{\partial}_J y^k)) +  H^l_z \Delta y^k \right)\nonumber\\
&= 
\epsilon_{j3l}\epsilon^{IJ} H^l_I \partial_J A^3_z  + \epsilon_{jkl}\left(H^l_z \Delta -\epsilon^{IJ} H^l_I  \partial_z \hat{\partial}_J \right)y^k  - \epsilon_{jkl}\epsilon^{IJ} H^l_I \partial_z \partial_J v^k +\epsilon_{j\alpha l}\epsilon^{IJ} H^l_I \partial_J A^\alpha_z
\end{align}
and
\begin{align}\label{C0}
\tilde{C}_0 &= B^I_j H^j_I + B^z_j H^j_z \nonumber\\
&=
\epsilon^{IJ} H^j_I \left(\partial_J A^j_z - \partial_z (\partial_J v^j + \hat{\partial}_J y^j)\right) + H^j_z \Delta y^j \nonumber \\
&=
\epsilon^{IJ} H^3_I \partial_J A^3_z + \left(H^j_z \Delta -\epsilon^{IJ} H^j_I \partial_z  \hat{\partial}_J \right) y^j -\epsilon^{IJ} H^j_I \partial_z \partial_J v^j + \epsilon^{IJ} H^\alpha_I \partial_J A^\alpha_z
\end{align}
where we have used $B^I_j = \epsilon^{IJ}\left(\partial_J A^j_z - \partial_z (\partial_J v^j + \hat{\partial}_J y^j) \right)$ and $B^z_j= \Delta y^j$. It can be read from (\ref{Cj}) and (\ref{C0}) that $h_I$ introduced in (\ref{3.1}) is $(\epsilon_{j\alpha l}\epsilon^{IJ} H^l_I \partial_J A^\alpha_z, \epsilon^{IJ} H^\alpha_I \partial_J A^\alpha_z)$. On the other hand, the gauge conditions (\ref{4.1}) are translated to these canonical variables as
\begin{equation}
G^j=v^j,\;\tilde{G}_j=x_j,\; \tilde{G}_0=E^z_3-f
\end{equation}
Recall that we assume $f$ to be independent of $z$.
The non-zero $H$'s evaluated at the gauge cut are
\begin{align*}
&H^1_x = f(1+\partial_y u^2)-E^z_2 (\partial_y u^3)\\
&H^2_x = E^z_1 (\partial_y u^3) - f(\partial_y u^1)\\
&H^3_x = E^z_2(\partial_y u^1) - E^z_1 (1+ \partial_y u^2)\\
&H^1_y = E^z_2 (\partial_x u^3) - f (\partial_x u^2)\\
&H^2_y = f (1+ \partial_x u^1) - E^z_1 (\partial_x u^3)\\
&H^3_y = E^z_1 (\partial_x u^2) - E^z_2 (1+ \partial_x u^1)\\
&H^3_z = (1+\partial_x u^1) (1+ \partial_y u^2)- (\partial_x u^2) (\partial_y u^1)
\end{align*}
and $H^\alpha_z = 0$.
The stability conditions for $\tilde{G}_j$ and $\tilde{G}_0$ at the gauge cut are
\begin{align}
&0=\{H, x_1\} = \partial^I \left[-\partial_I (\lambda^2 H^3_z)- \partial_z (\lambda^3 H^2_I - \lambda^2 H^3_I + \lambda H^1_I)\right]\label{stability of x1}\\
&0=\{H, x_2\} = \partial^I \left[\partial_I (\lambda^1 H^3_z)- \partial_z (\lambda^1 H^3_I - \lambda^3 H^1_I + \lambda H^2_I)\right]\label{stability of x2}\\
&0=\{H, x_3\} = \partial^I \left[\partial_I (\lambda H^3_z)- \partial_z (\lambda^2 H^1_I - \lambda^1 H^2_I + \lambda H^3_I)\right]\label{stability of x3}\\
&0=\{H, E^z_3\} = \hat{\partial}^I \left[\lambda^2 H^1_I - \lambda^1 H^2_I + \lambda H^3_I \right] \label{stability of Ez3}
\end{align}
respectively, which are supposed to be solved for $\lambda^i$ and $\lambda$. From (\ref{stability of x3}) and (\ref{stability of Ez3}), one concludes 
\begin{align}
&\partial_I (\lambda H^3_z)- \partial_z (\lambda^2 H^1_I - \lambda^1 H^2_I + \lambda H^3_I)= \hat{\partial}_I g_1 \label{x3 1}\\
& \lambda^2 H^1_I - \lambda^1 H^2_I + \lambda H^3_I= \partial_I g_2 \label{Ez3 1}
\end{align}
respectively, where $g_1$ and $g_2$ are certain 0-forms. Inserting (\ref{Ez3 1}) into (\ref{x3 1}) and applying $\hat{\partial}^I$ on both sides lead to $\Delta g_1 = 0$. Since $g_1$ is harmonic and also decaying at infinity due to the boundary conditions, then $g_1=0$. Hence (\ref{x3 1}) turns to 
\begin{equation}\label{equations for N1 and N2}
\lambda^2 H^1_I - \lambda^1 H^2_I = \partial_I \partial_z^{-1}(\lambda H^3_z)  + \lambda H^3_I + g_I (x,y)
\end{equation}
in which $g_I$ are functions depending only on $x, y$. Going to infinity along $z$-axis while $x, y$ are finite and fixed shows that $g_I = \epsilon_{IJ} \delta^J_\alpha \lambda^\alpha_0 $ where $\lambda^\alpha_0$ are the leading terms of $\lambda^\alpha$ (recall that $\lambda^i$ are of $O(1)$ and the leading terms are constants). Now, (\ref{equations for N1 and N2}) are two algebraic equations and can be solved for $\lambda^\alpha$ as 
\begin{equation}\label{lambda alpha1}
\lambda^\alpha =\frac{1}{H^3_z}\left[\epsilon^{IJ} H^\alpha_I \left(\partial_J\partial_z^{-1}(\lambda H^3_z)-H^3_J \right) + \lambda_0^\alpha \right]
\end{equation}
Notice that from plugging (\ref{lambda alpha1}) in (\ref{stability of x1}) and (\ref{stability of x2}), a system of two integro-differential equations arises which is too complicated to be solved. Even if one could solve the system for $\lambda^3$ and $\lambda$, the resulting physical Hamiltonian would be very complicated to be quantized. Hence, we leave further analysis of this gauge and in the $(A,E)$ description, we will continue the quantization process with MLA-ETTA gauge which led to a relatively simple physical Hamiltonian (\ref{4.24}).

\subsubsection{ETA-LEA gauge in $(B,f)$ description}
\label{s4.2.3}
First, recall that in this gauge condition we have $z=\{(f_I^i, B^I_i), (f^3_z, B^z_3)\}$ and $r=(f^\alpha_z, B^z_\alpha)$.
The equations of the stability of the gauge conditions are
\begin{align}\label{Stability equations 1}
\dot{\sigma}^i_I=\{H, f^i_I (y)\}=& \int d^3x \; [\lambda^k \partial_a + \tilde{\lambda}^j \epsilon_{jkl}H^l_a + \bar{\lambda} H^k_a] \{B^a_k (x), f^i_I (y)\}\nonumber\\
=&
- \partial_I \lambda^i + \tilde{\lambda}^j \epsilon_{jil}H^l_I + \bar{\lambda} H^i_I\nonumber\\
\dot{\sigma}=\{H, f^3_z (y)\}=& \int d^3x \; [\lambda^k \partial_a + \tilde{\lambda}^j \epsilon_{jkl}H^l_a + \bar{\lambda} H^k_a] \{B^a_k (x), f^3_z (y)\}\nonumber\\
=&
- \partial_z \lambda^3 + \tilde{\lambda}^j \epsilon_{j3l}H^l_z + \bar{\lambda} H^3_z
\end{align}
which are supposed to be uniquely solved for the Lagrange multipliers. As it is mentioned before, it is sufficient to solve them at the gauge cut where the system of equations (\ref{Stability equations 1}) can be represented as 
\begin{equation}\label{System 1}
\left[ {\begin{array}{ccccccc}
   -\partial_x & 0 & 0 & 0 & 0 & H^2_x & H^1_x \\
   0 & -\partial_x & 0 & 0 & 0 & -H^1_x & H^2_x \\
   0 & 0 & -\partial_x & -H^2_x & H^1_x & 0 & 0 \\
   -\partial_y & 0 & 0 & 0 & 0 & H^2_y & H^1_y \\
   0 & -\partial_y & 0 & 0 & 0 & -H^1_y & H^2_y \\
   0 & 0 & -\partial_y & -H^2_y & H^1_y & 0 & 0 \\
   0 & 0 & -\partial_z & 0 & 0 & 0 & H^3_z \\
  \end{array} } \right]
  \left[ {\begin{array}{c}
   \lambda^1  \\
   \lambda^2  \\
   \lambda^3  \\
   \tilde{\lambda}^1  \\
   \tilde{\lambda}^2  \\
   \tilde{\lambda}^3  \\
   \bar{\lambda}  \\
  \end{array} } \right]
  =
  \left[ {\begin{array}{c}
   \dot{\sigma}^1_x  \\
   \dot{\sigma}^2_x  \\
   \dot{\sigma}^3_x  \\
   \dot{\sigma}^1_y  \\
   \dot{\sigma}^2_y  \\
   \dot{\sigma}^3_y  \\
   \dot{\sigma}  \\
  \end{array} } \right]=: \dot{\Sigma}^I
\end{equation}
\\
First, the space of the functions needs to be determined in such a way that all integral constants are fixed while solving the system of equations. For this purpose,  we will work only with functions that are of the following form
\begin{align}\label{asym.behaviour of Lagrange mul.}
\lambda^i &= \lambda^i_0 \delta_{ij} \delta^j_a x^a + O(r^{-1})\nonumber  \\
\tilde{\lambda}^i &= \tilde{\lambda}^i_0 + O(r^{-1})\nonumber \\
\bar{\lambda} &= \bar{\lambda}_0 + O(r^{-2})
\end{align}
where $\lambda^i_0, \tilde{\lambda}^i_0, \bar{\lambda}_0$ are arbitrary constants.
Note that the first equation of (\ref{asym.behaviour of Lagrange mul.}) is completely consistent with (\ref{BC on lambda}) and (\ref{PC on lambda}) and in the second and the third equation rotations and boosts are excluded, respectively, as there are not well-defined generators for them \cite{11}. The reason for the lack of $r^{-1}$ term in the lapse function is that in the following calculations, when one uses anti-derivatives, it would lead to a logarithmic divergence preventing us from specifying some integration constants. This is exactly why we also use 
\begin{equation}\label{Asymp f new}
f_a^j \to c^j_a  + O(1/r) 
\end{equation} 
instead of (\ref{Asymp f}) in what follows. \\
\\
Here, we wish to work with $\dot{\Sigma}_I = 0$. Solving the first equation of (\ref{System 1}) for $\lambda^1$ results in $\lambda^1= \bar{\lambda}_0 x + \partial_x^{-1}\left(H^1_x \bar{\lambda}+ H^2_x \tilde{\lambda}^3\right)+ g_1(y,z)$ where we have used $H^1_x = 1 + O(r^{-2})$ and $\bar{\lambda}=\bar{\lambda}_0 + O(r^{-2})$ and the fact that all constants are in the kernel of $\partial_a^{-1}$. Noting that based on (\ref{asym.behaviour of Lagrange mul.}), $\lambda^1 - \bar{\lambda}_0 x\to 0$ asymptotically and moving to infinity along the $x$-axis at fixed finite values of $y,z$, one observes  $g_1=0$. Hence,
\begin{equation}\label{equation for lambda1}
\lambda^1= \bar{\lambda}_0 x + \partial_x^{-1}\left(H^1_x \bar{\lambda}+ H^2_x \tilde{\lambda}^3\right)
\end{equation}
The same argument can be employed to solve the fifth equation of (\ref{System 1}) for $\lambda^2$ as 
$\lambda^2= \bar{\lambda}_0 y + \partial_x^{-1}\left(H^2_y \bar{\lambda}- H^1_y \tilde{\lambda}^3\right)+ g_2(x,z)$. Going to infinity along $y$-axis at fixed finite values of $x,z$ results in $g_2=0$, because $\lambda^2 - \bar{\lambda}_0 y\to 0$. Therefore,
\begin{equation}\label{equation for lambda2}
\lambda^2= \bar{\lambda}_0 y + \partial_x^{-1}\left(H^2_y \bar{\lambda}- H^1_y \tilde{\lambda}^3\right)
\end{equation}
The last equation can be solved for $\lambda^3$ as $\lambda^3= \bar{\lambda}_0 z + \partial_z^{-1}\left(H^3_z \bar{\lambda}\right)+g_3(x,y)$ where we used $H^3_z = 1 + O(r^{-2})$. Again, going to infinity along the $z$-axis while $x,y$ are fixed and finite, we deduce $g_3=0$. Consequently,
\begin{equation}\label{equation for lambda3}
\lambda^3= \bar{\lambda}_0 z + \partial_z^{-1}\left(H^3_z \bar{\lambda}\right)
\end{equation}
Plugging $\lambda^1$ and $\lambda^2$ into the second and fourth equations gives us a system of two integro-differential equations for $\tilde{\lambda}^3,\bar{\lambda}$  
\begin{align}
&(-\partial_x \partial_y^{-1}H^2_y + H^2_x)\bar{\lambda}-(-\partial_x \partial_y^{-1}H^1_y + H^1_x)\tilde{\lambda}^3 =0\label{integro-diff 3}\\
&(-\partial_y \partial_x^{-1}H^1_x + H^1_y)\bar{\lambda} +(-\partial_y \partial_x^{-1}H^2_x + H^2_y)\tilde{\lambda}^3 =0\label{integro-diff 4}
\end{align}
Looking at (\ref{Hs in ETA-LEA}), one can easily check that all constants belong to the kernel of both operator $Y_2 := -\partial_x \partial_y^{-1}H^2_y + H^2_x$ and $X_1:= -\partial_y \partial_x^{-1}H^1_x + H^1_y$. In fact, if $u$ is a constant 
\begin{align*}
Y_2 u =& -\partial_x \partial_y^{-1}(H^2_y u) + H^2_x u\\
=&
\; u \left(-\partial_x \partial_y^{-1}H^2_y + H^2_x \right)\\
=&
\; u \left(-\partial_x \partial_y^{-1}(1+\partial_y f^1_z) + \partial_x f^1_z \right)\\
=& \; 0
\end{align*}
and by a similar argument $X_1 u = 0$ as well.
On the other hand, both operators $Y_1 := -\partial_x \partial_y^{-1}H^1_y + H^1_x$ and $X_2:= -\partial_y \partial_x^{-1}H^2_x + H^2_y$ acts on constants like the identity, in the sense that $Y_1 u = X_2 u = u$ and the reason would be
\begin{align*}
Y_1 u =& -\partial_x \partial_y^{-1}(H^1_y u) + H^1_x u\\
=&
\; u \left(-\partial_x \partial_y^{-1}H^1_y + H^1_x \right)\\
=&
\; u \left(-\partial_x \partial_y^{-1}(-\partial_y f^2_z) + (1-\partial_x f^2_z) \right)\\
=& \; u
\end{align*}
and by the same reasoning $X_2 u= u$, for all $u = constant$.
 Thus, two integro-differential equations (\ref{integro-diff 3}) and (\ref{integro-diff 4}) are equivalent to 
\begin{align}
&Y_2 (\bar{\lambda}-\bar{\lambda}_0)- Y_1 (\tilde{\lambda}^3 -\tilde{\lambda}^3_0) -\tilde{\lambda}^3_0 =0\label{integro-diff 5}\\
&X_1 (\bar{\lambda}-\bar{\lambda}_0) + X_2 (\tilde{\lambda}^3 -\tilde{\lambda}^3_0) + \tilde{\lambda}^3_0 =0\label{integro-diff 6}
\end{align}
Since $\bar{\lambda}-\bar{\lambda}_0$ and $\tilde{\lambda}^3 -\tilde{\lambda}^3_0$ are of $O(r^{-1})$, the highest order term of both (\ref{integro-diff 5}) and (\ref{integro-diff 6}) is $\tilde{\lambda}^3_0$ which must be vanish separately. Hence, $\tilde{\lambda}^3= \sum_{n=1}^\infty \tilde{\lambda}^3_n r^{-n}$ and (\ref{integro-diff 5}) and (\ref{integro-diff 6}) reduce to
\begin{align}
&Y_2 (\bar{\lambda}-\bar{\lambda}_0)- Y_1 \tilde{\lambda}^3 =0\label{integro-diff 7}\\
&X_1 (\bar{\lambda}-\bar{\lambda}_0) + X_2 \tilde{\lambda}^3=0\label{integro-diff 8}
\end{align}
It follows from (\ref{integro-diff 7}) that $ \tilde{\lambda}^3 = Y_1^{-1} Y_2 (\bar{\lambda}- \bar{\lambda}_0)+ \tilde{\kappa}$, where $\tilde{\kappa}$ is in the kernel of $Y_1$. Since $\tilde{\lambda}^3 = O(r^{-1})$, $\tilde{\kappa}$ has to be of the form $\tilde{\kappa}= \sum_{n=1}^\infty \tilde{\kappa}_n r^{-n}$ where $\tilde{\kappa}_n$ are functions on the asymptotic sphere. We have
\begin{align}\label{Y1 kappa}
0= Y_1 \tilde{\kappa}= H^1_x \tilde{\kappa} -  \partial_y \partial_x^{-1} (H^1_y \tilde{\kappa})
\end{align}
Since $H^1_I= \delta^1_I +  O(r^{-2})$ and $\tilde{\kappa}= O(r^{-1})$, the first and the second terms of (\ref{Y1 kappa}) are of $O(r^{-1})$ and $O(r^{-3})$, respectively. Thus, $\tilde{\kappa}_1/r$ is the highest order term existing in (\ref{Y1 kappa}) that has to vanish separately, i.e. $\tilde{\kappa}_1 = 0$ and consequently $\tilde{\kappa}= O(r^{-2})$. Now, the highest order term in (\ref{Y1 kappa}) is $\tilde{\kappa}_2/r^2$ that has to vanish by the same reasoning. By induction one concludes $\tilde{\kappa}_n =0$ for all $n>0$ which means $\tilde{\kappa}=0$.
Therefore,
\begin{equation}\label{lambda3 in terms of lambda}
\tilde{\lambda}^3 = Y_1^{-1} Y_2 (\bar{\lambda}- \bar{\lambda}_0)
\end{equation}
Plugging (\ref{lambda3 in terms of lambda}) into (\ref{integro-diff 8}), we have
\begin{equation}\label{equation for lambda bar}
0=(X_2 Y_1^{-1}Y_2 + X_1 ) (\bar{\lambda}- \bar{\lambda}_0) = -\partial_x^{-1}\partial_y [Y_1 + Y_2 Y_1^{-1} Y_2](\bar{\lambda}- \bar{\lambda}_0)
\end{equation} 
where in writing the second equality we used $X_\alpha =-\partial_x^{-1}\partial_y Y_\alpha $.
(\ref{equation for lambda bar}) tells us that $\partial_y [Y_1 + Y_2 Y_1^{-1} Y_2](\bar{\lambda}- \bar{\lambda}_0)=g(y,x)$ in which $g$ is an arbitrary function depending only on $y , z$. Since $\bar{\lambda}- \bar{\lambda}_0 = O(r^{-1})$, moving to infinity along the $x$-axis at fixed finite values of $y, z$ shows that $g(y,x)=0$. Therefore, $[Y_1 + Y_2 Y_1^{-1} Y_2](\bar{\lambda}- \bar{\lambda}_0)=h(x,z)$ where $h$ is an arbitrary function not depending on $y$. Again, going to infinity along the $y$-axis while $x, z$ have fixed finite values results in $h(x, z)=0$.
 Thus, (\ref{equation for lambda bar}) is equivalent to
\begin{equation}\label{simpler equation for lambda bar}
[Y_1 + Y_2 Y_1^{-1} Y_2](\bar{\lambda}- \bar{\lambda}_0)=0
\end{equation} 
In general, it is easy to show that $(S+P)^{-1}= S^{-1}- S^{-1} P (S+P)^{-1}$ for every two operators $S, P$.
By repeatedly inserting this relation into its r.h.s, one obtains
\begin{equation}\label{inverse of sum}
(S+P)^{-1}= S^{-1} \sum_{n=0}^{\infty} (-P S^{-1})^n 
\end{equation} 
Based on this relation, one gets
\begin{equation}\label{inverse of Y2}
Y_1^{-1}= (H^1_x -\partial_y \partial_x^{-1}H^1_y)^{-1} = \frac{1}{H^1_x} \sum_{n=0}^{\infty} (\partial_y \partial_x^{-1}\frac{ H^1_y}{H^1_x})^n 
\end{equation}
Since $H^\alpha_I = \delta^\alpha_I + O(r^{-2})$, $Y_1^{-1}$ is expanded as $Y_1^{-1}=1+ O(r^{-2})$ and $Y_2= - \partial_x \partial_y^{-1} + O(r^{-2})$. Assuming $\bar{\lambda}- \bar{\lambda}_0 = \sum_{n=1}^\infty \bar{\lambda}_n r^{-n}$ in which $\bar{\lambda}_n= \bar{\lambda}_n (\theta, \varphi)$ in the spherical coordinates, we can extract the highest order term of (\ref{simpler equation for lambda bar}) as $\left( \partial^2_x \partial^{-2}_y + 1\right) \frac{\bar{\lambda}_1}{r}=0$. By applying $\partial_y^2$ on both sides of this equation, we see that $\bar{\lambda}_1/r$ has to satisfy the 2-dimensional Laplace's equation, that is
\begin{equation}
(\partial_x^2 + \partial_y^2) \frac{\tilde{\lambda}_1}{r}=0
\end{equation}
If one defines $R=\sqrt{x^2+y^2}$, it is easy to rewrite the Laplace's equation in the polar coordinate system in $x-y$ plane
\begin{equation}\label{Laplace in polar coordinates}
\left(\frac{\partial^2}{\partial R^2}+ \frac{1}{R} \frac{\partial}{\partial R} + \frac{1}{R^2} \frac{\partial^2}{\partial \varphi^2}\right) \frac{\bar{\lambda}_1(\theta, \varphi)}{\sqrt{R^2 + z^2}}=0
\end{equation}
where $z= r \cos \theta$ and $r= \sqrt{R^2 + z^2}$. By carrying out the derivatives, (\ref{Laplace in polar coordinates}) is reduced to 
\begin{equation}\label{simpler Laplace in polar coordinates}
-\bar{\lambda}_1 \left(\frac{2z^2 - R^2}{(R^2 + z^2)^2} \right)+ \frac{1}{R^2} \partial_\varphi^2 \bar{\lambda}_1 =0
\end{equation}
Moving to infinity along $z$-axis at fixed finite values of $x, y$ results in $\partial_\varphi^2 \bar{\lambda}_1=0$. Inserting this into (\ref{simpler Laplace in polar coordinates}) shows that $\bar{\lambda}_1$ has to vanish. Thus, $\bar{\lambda}- \bar{\lambda}_0 = \sum_{n=2}^\infty \bar{\lambda}_n r^{-n}$. Now the highest order term of (\ref{simpler equation for lambda bar}) is $\left( \partial^2_y \partial^{-2}_x + 1\right) \frac{\bar{\lambda}_2}{r^2}=0$ and by a similar argument it is easily concluded that $\bar{\lambda}_2=0$ and finally by induction $\bar{\lambda}_n=0$ for all $n>0$. As a final result of this part, we have $\bar{\lambda}- \bar{\lambda}_0 =0$ and consequently $\tilde{\lambda}^3= 0$ (recall (\ref{lambda3 in terms of lambda})). It follows from (\ref{equation for lambda3}) that $\lambda^3 = \bar{\lambda}_0 z + \bar{\lambda}_0 \partial_z^{-1} H^3_z$.
Accordingly, the third and sixth equations of (\ref{System 1}) are a system of two algebraic equations that simply results in
\begin{align}
\tilde{\lambda}^\alpha = \frac{ \bar{\lambda}_0}{H^3_z}(H^\alpha_y \partial_x - H^\alpha_x \partial_y) \partial_z^{-1} H^3_z 
\end{align}
for $\alpha=1,2$ because $H^1_x H^2_y-H^2_x H^1_y = H^3_z \neq 0$. Finally, (\ref{equation for lambda1}) and (\ref{equation for lambda2}) lead to
\begin{align}
\lambda^1=&\; \bar{\lambda}_0 x + \bar{\lambda}_0 \partial_x^{-1} H^1_x = \bar{\lambda}_0 x - \bar{\lambda}_0 f^2_z \nonumber\\
\lambda^2=&\;  \bar{\lambda}_0 y + \bar{\lambda}_0 \partial_x^{-1}H^2_y  = \bar{\lambda}_0 y + \bar{\lambda}_0 f^1_z
\end{align}
respectively. \\
This ends proving that the solution of the system of PDEs (\ref{System 1}) with $\dot{\Sigma}^I =0$ is of the form 
\begin{equation}\label{Lambda 1}
\Lambda^I_0 =\bar{\lambda}_0 \left(x -  f^2_z, y + f^1_z, z + \partial_z^{-1} H^3_z, \frac{1}{H^3_z}(H^1_y \partial_x - H^1_x \partial_y) \partial_z^{-1} H^3_z, \frac{1}{H^3_z}(H^2_y \partial_x - H^2_x \partial_y) \partial_z^{-1} H^3_z, 0, 1 \right)^T
\end{equation}
where $\bar{\lambda}_0 $ is an arbitrary constant. In order to have a unique solution, it is required to fix the asymptotic behaviour of $\bar{\lambda}$. For simplicity, we consider $\bar{\lambda}_0 = 1$ which means that from the scratch one is supposed to work only with those lapse functions in (\ref{asym.behaviour of Lagrange mul.}) which are of the form $\bar{\lambda}=1+O(r^{-2})$.
Based on the theorem (\ref{th3.2}), to obtain the corresponding physical Hamiltonian for this gauge fixing, it is sufficient to multiply $\Lambda_0^I$ to $(h_I)_{G=0}=(\partial_z B^z_1, \partial_z B^z_2, 0, B^z_2 H^3_z, -B^z_1 H^3_z, 0, 0)$. Consequently, for this special gauge fixing, one achieves
\begin{align}\label{Phys. Ham. 3}
h_{\Sigma^I =0}=& \int d^3x\;  [ (\delta^\beta_I x^I + \epsilon^{\alpha \beta} f^\alpha_z) (\partial_z B^z_\beta)+ \epsilon^{\alpha \beta}\epsilon^{IJ}B^z_\alpha  H^\beta_I \partial_J \partial_z^{-1}H^3_z]\nonumber \\
=& \int d^3x\; \epsilon^{\alpha \beta} [ f^\alpha_z (\partial_z B^z_\beta)+ \epsilon^{IJ}B^z_\alpha  H^\beta_I \partial_J \partial_z^{-1}H^3_z]  + \oint dS_a \; \delta^\beta_I x^I  \delta_z^a B^z_\beta \nonumber \\
=& \int d^3x\; \epsilon^{\alpha \beta} [ f^\alpha_z (\partial_z B^z_\beta)+ \epsilon^{IJ}B^z_\alpha  H^\beta_I \partial_J \partial_z^{-1}H^3_z]
\end{align}
where the surface term has been dropped because $B^z_\beta $ is $O(r^{-3})$ odd.\\
Two other appropriate choices for gauge fixing turn out to be $\Sigma^I =(0, 0, \tau, 0, 0, 0 , 0 )^T$ and $\Sigma^I =(0, 0, 0, 0, 0, \tau , 0 )^T$ which lead to 
\begin{align}
\Lambda^I_0 &= \left(x -  f^2_z, y + f^1_z, z + \partial_z^{-1} H^3_z, \frac{H^1_y}{H^3_z}+ \frac{1}{H^3_z}(H^1_y \partial_x - H^1_x \partial_y) \partial_z^{-1}H^3_z, \frac{H^2_y}{H^3_z}+\frac{1}{H^3_z}(H^2_y \partial_x - H^2_x \partial_y) \partial_z^{-1}H^3_z, 0, 1\right),\\
\Lambda^I_0 &= \left(x -  f^2_z, y + f^1_z, z + \partial_z^{-1} H^3_z, \frac{1}{H^3_z}(H^1_y \partial_x - H^1_x \partial_y) \partial_z^{-1}H^3_z -\frac{H^1_x}{H^3_z}, \frac{1}{H^3_z}(H^2_y \partial_x - H^2_x \partial_y) \partial_z^{-1}H^3_z -\frac{H^2_x}{H^3_z}, 0, 1\right)
\end{align}
respectively. And the corresponding physical Hamiltonian are obtained as 
\begin{align}\label{Phys. Ham. 1}
h_{\Sigma^I =(0, 0, \tau, 0, 0, 0 , 0 )^T}=& \int d^3x\; \epsilon^{\alpha \beta} [f^\alpha_z (\partial_z B^z_\beta)+ \epsilon^{IJ}B^z_\alpha  H^\beta_I \partial_J \partial_z^{-1}H^3_z + H^\alpha_y B^z_\beta]
\end{align}
and
\begin{align}\label{Phys. Ham. 2}
h_{\Sigma^I =(0, 0, 0, 0, 0, \tau , 0 )^T}=& \int d^3x\; \epsilon^{\alpha \beta} [f^\alpha_z (\partial_z B^z_\beta)+ \epsilon^{IJ}B^z_\alpha  H^\beta_I \partial_J \partial_z^{-1}H^3_z - H^\alpha_x B^z_\beta]
\end{align}
respectively, where the expressions of $H^\alpha_I$ and $H^3_z$ in terms of $f^\alpha_z$ have been written in (\ref{Hs in ETA-LEA}). \\
The Physical Hamiltonians (\ref{Phys. Ham. 3}), (\ref{Phys. Ham. 1}) and (\ref{Phys. Ham. 2}) have the following features:\\
I. Linearity in momentum $B^z_\alpha$.\\
II. Polynomiality in the configuration variable $f^\alpha_z$.\\
III. Spatial non-locality.\\
Since three physical Hamiltonians obtained here are very similar, in what follows we will only work with (\ref{Phys. Ham. 3}). 
\\
\\
At the end of this subsection, we derive the equations of motion using the physical Hamiltonian (\ref{Phys. Ham. 3}) and $\{B^z_\alpha(x),f^\beta_z(y)\}=\delta^\alpha_\beta \delta(x,y)$. For $f^\alpha_z$, one can effortlessly see that
\begin{align}\label{EOM f}
\dot{f}^\alpha_z(x)=\{h, f^\alpha_z(x)\}= \epsilon^{\alpha \beta} [\partial_z f^\beta_z + \epsilon^{IJ} H^\beta_I \partial_J \partial_z^{-1}H^3_z]
\end{align}
in which
\begin{align}
&H^\alpha_I = \delta^\alpha_I - \epsilon^{\alpha \beta} \partial_I f^\beta_z\\
&H^3_z= 1 + \epsilon^{\alpha \beta} \delta^I_\beta \partial_I f^\alpha_z +\frac{1}{2} \epsilon^{\alpha \beta} \epsilon^{IJ} (\partial_I f^\alpha_z)(\partial_J f^\beta_z)
\end{align}
Unsurprisingly, the equation of motion for $f^\alpha_z$ closes on itself thanks to the linearity feature of the physical Hamiltonian in $B^z_\alpha$. Therefore, (\ref{EOM f}) can be solved separately without the need to refer to the equation of motion of $B^a_i$.
To obtain the time evolution $B^a_i$, one requires to know the variations of $H^\beta_I$ and $H^3_z$ with respect to $f^\alpha_z$ which are straightforwardly derived $\frac{\delta H^\beta_I (x)}{\delta f^\alpha_z (y)}= \epsilon^{\alpha \beta} \partial_I \delta (x,y)$ and $\frac{\delta H^3_z (x)}{\delta f^\alpha_z (y)}= \epsilon^{\alpha \beta} \left[\delta^I_\beta + \epsilon^{IJ} (\partial_J f^\beta_z) \right]\partial_I \delta (x,y)$. Employing these equations, we get
\begin{align}\label{EOM B}
\dot{B}_\alpha^z(x)=&\{h, B_\alpha^z(x)\}= -\epsilon^{\alpha \beta} (\partial_z B^z_\beta)\nonumber\\
 &- \int d^3x\; \epsilon^{\gamma \beta} \left\{ \epsilon^{IJ}B^z_\gamma (\partial_J \partial_z^{-1}H^3_z) \epsilon^{\alpha \beta}\partial_I \delta (x,y) + \epsilon^{KJ}B^z_\gamma  H^\beta_K \partial_J \partial_z^{-1}\left(\epsilon^{\alpha \lambda} \left[\delta^I_\lambda + \epsilon^{IL} (\partial_L f^\lambda_z) \right]\partial_I \delta (x,y) \right)\right\}\nonumber\\
 =&\; -\epsilon^{\alpha \beta} (\partial_z B^z_\beta)
 + \epsilon^{IJ}(\partial_I B^z_\alpha) (\partial_J \partial_z^{-1}H^3_z) 
 - \int d^3x\; \epsilon^{\gamma \beta} \epsilon^{KJ} \epsilon^{\alpha \lambda} \left(\delta^I_\lambda + \epsilon^{IL} (\partial_L f^\lambda_z) \right) \partial_J \partial_z^{-1}(B^z_\gamma  H^\beta_K) \partial_I \delta (x,y)\nonumber\\
 =&\;-\epsilon^{\alpha \beta} (\partial_z B^z_\beta)
 + \epsilon^{IJ}(\partial_I B^z_\alpha) (\partial_J \partial_z^{-1}H^3_z) 
 +\epsilon^{\gamma \beta} \epsilon^{KJ} \epsilon^{\alpha \lambda}\left(\delta^I_\lambda + \epsilon^{IL} (\partial_L f^\lambda_z) \right)\partial_I \partial_J \partial_z^{-1}(B^z_\gamma  H^\beta_K) 
\end{align}
where in the last step we used $\partial_I \left[\delta^I_\lambda + \epsilon^{IL} (\partial_L f^\lambda_z) \right]=0$.
As the equations (\ref{EOM f}) and (\ref{EOM B}) are complicated to be solved, we leave the further analysis of them for future work.
\subsubsection{ETTA-ELA gauge in $(B,f)$ description}
\label{s4.2.4}
First, recall that in this gauge condition we have $z=\{(f_I^\alpha, B^I_\alpha), (f^j_z, B^z_j)\}$ and $r=(f^3_I, B^I_3)$.
The equations of the stability of the gauge conditions are
\begin{align}\label{Stability equations 2}
\dot{\sigma}^\alpha_I=\{H, f^\alpha_I (y)\}=& \int d^3x \; [\lambda^k \partial_a + \tilde{\lambda}^j \epsilon_{jkl}H^l_a + \bar{\lambda} H^k_a] \{B^a_k (x), f^\alpha_I (y)\}\nonumber\\
=&
- \partial_I \lambda^\alpha + \tilde{\lambda}^j \epsilon_{j\alpha l}H^l_I + \bar{\lambda} H^\alpha_I\nonumber\\
\dot{\sigma}^i=\{H, f^i_z (y)\}=& \int d^3x \; [\lambda^k \partial_a + \tilde{\lambda}^j \epsilon_{jkl}H^l_a + \bar{\lambda} H^k_a] \{B^a_k (x), f^i_z (y)\}\nonumber\\
=&
- \partial_z \lambda^i + \tilde{\lambda}^j \epsilon_{jil}H^l_z + \bar{\lambda} H^i_z
\end{align}
which are supposed to be uniquely solved for the Lagrange multipliers. The system of equations (\ref{Stability equations 2}) can be represented at the gauge cut as 
\begin{equation}\label{System 2}
\left[ {\begin{array}{ccccccc}
   -\partial_x & 0 & 0 & 0 & 0 & 0 & H^1_x \\
   0 & -\partial_x & 0 & 0 & 0 & -H^1_x & 0 \\
   -\partial_y & 0 & 0 & 0 & 0 & H^1_x & 0 \\
   0 & -\partial_y & 0 & 0 & 0 & 0 & H^1_x \\
   -\partial_z & 0 & 0 & 0 & -1 & H^2_z & H^1_z \\
   0 & -\partial_z & 0 & 1 & 0 & -H^1_z & H^2_z \\
   0 & 0 & -\partial_z & -H^2_z & H^1_z & 0 & 1 \\
  \end{array} } \right]
  \left[ {\begin{array}{c}
   \lambda^1  \\
   \lambda^2  \\
   \lambda^3  \\
   \tilde{\lambda}^1  \\
   \tilde{\lambda}^2  \\
   \tilde{\lambda}^3  \\
   \bar{\lambda}  \\
  \end{array} } \right]
  =
  \left[ {\begin{array}{c}
   \dot{\sigma}^1_x  \\
   \dot{\sigma}^2_x  \\
   \dot{\sigma}^3_x  \\
   \dot{\sigma}^1_y  \\
   \dot{\sigma}^2_y  \\
   \dot{\sigma}^3_y  \\
   \dot{\sigma}  \\
  \end{array} } \right]=: \dot{\Sigma}^I
\end{equation}
\\
Again,  we will work with the space of functions introduced in (\ref{asym.behaviour of Lagrange mul.}). And we consider (\ref{Asymp f new}), (\ref{Asymp B}) and (\ref{New PC}) as boundary conditions imposed on the canonical variables.\\
\\ 
Now, we can solve (\ref{System 2}) with the assumption $\dot{\Sigma}^I=0$. From the first and the fourth equations of (\ref{System 2}), it is immediately concluded that $\lambda^\alpha= \lambda_0 \delta^\alpha_I x^I + \delta_\alpha^I \partial_I^{-1}(\bar{\lambda}H^1_x)+g_\alpha(y,z)$. Since it is assumed that $H^1_x = 1+ O(r^{-2})$, the asymptotic behaviours of (\ref{asym.behaviour of Lagrange mul.}) shows that $g_1 = 0$. Thus,
\begin{equation}\label{lambda alpha}
\lambda^\alpha= \lambda_0 \delta^\alpha_I x^I + \delta_\alpha^I \partial_I^{-1}(\bar{\lambda}H^1_x)
\end{equation}
Plugging (\ref{lambda alpha}) into the second and the third equations of (\ref{System 2}) gives us 
\begin{align}\label{the second one}
-\partial_x \partial_y^{-1} (\bar{\lambda}H^1_x) - H^1_x \tilde{\lambda}^3=0\nonumber\\ 
-\partial_y \partial_x^{-1} (\bar{\lambda}H^1_x) + H^1_x \tilde{\lambda}^3=0
\end{align}
respectively. One can easily solve the first equation of (\ref{the second one}) for $\tilde{\lambda}^3$ and gets
\begin{equation}\label{lambda tilde 3}
\tilde{\lambda}^3= -\frac{1}{H^1_x}\partial_x \partial_y^{-1} (\bar{\lambda}H^1_x)
\end{equation}
Inserting (\ref{lambda tilde 3}) into the second equation of (\ref{the second one}) results in $(\partial_y \partial_x^{-1}+\partial_x \partial_y^{-1}) u =0$ where $u:= \bar{\lambda}H^1_x= \bar{\lambda}_0 + O(r^{-2})$. If one applies the operator $\partial_x \partial_y$ on both sides of this equation, one sees that $u$ has to satisfy the 2-dimensional Laplace's equation $0=(\partial_x^2 + \partial_y^2)u = (\partial_x^2 + \partial_y^2)(u-\bar{\lambda}_0) $ in which $u-\bar{\lambda}_0$ is of $O(r^{-2})$ and therefore it can be written as $u-\bar{\lambda}_0 = \sum_{n=2}^\infty u_n r^{-n} $. The highest order term of the Laplace's equation under consideration is $(\partial_x^2 + \partial_y^2) (u_2 r^{-2})$ which can be rewritten in the polar coordinate system in $x-y$ plane as 
\begin{equation}\label{Laplace in polar coordinates 2}
\left(\frac{\partial^2}{\partial R^2}+ \frac{1}{R} \frac{\partial}{\partial R} + \frac{1}{R^2} \frac{\partial^2}{\partial \varphi^2}\right) \frac{u_2(\theta, \varphi)}{R^2 + z^2}=0
\end{equation}
where $R=\sqrt{x^2+y^2}$, $z= r \cos \theta$ and $r= \sqrt{R^2 + z^2}$. By carrying out the derivatives, (\ref{Laplace in polar coordinates}) is reduced to 
\begin{equation}\label{simpler Laplace in polar coordinates 2}
2 u_2 \left(\frac{R^2-z^2}{(R^2 + z^2)^2} \right)+ \frac{1}{R^2} \partial_\varphi^2 u_2 =0
\end{equation}
Moving to infinity along $z$-axis at fixed finite values of $x, y$ results in $\partial_\varphi^2 u_2=0$. Inserting this into (\ref{simpler Laplace in polar coordinates 2}) shows that $u_2$ has to vanish. Thus, $u-\bar{\lambda}_0= \sum_{n=3}^\infty \tilde{\lambda}_n r^{-n}$. By repeating the same argument for the lowest order term of the Laplace's equation which is now $(\partial_x^2 + \partial_y^2) (u_3 r^{-3})$, we see $u_3 =0$ and finally by induction $u_n=0$ for all $n>1$ which means $\bar{\lambda}H^1_x=u=\bar{\lambda}_0$. Consequently, from (\ref{lambda alpha}) and (\ref{lambda tilde 3}), it follows that
\begin{equation}\label{Final lambda alpha and tilde}
\lambda^\alpha= \lambda_0 \delta^\alpha_I x^I, \;\;\;\;  \tilde{\lambda}^3=0
\end{equation}
respectively. Plugging (\ref{Final lambda alpha and tilde}) and $\bar{\lambda}= \frac{\bar{\lambda}_0}{H^1_x}$ into the fifth and the sixth equations of (\ref{System 2}), we obtain
\begin{equation}
\tilde{\lambda}^\alpha= -\frac{\bar{\lambda}_0}{H^1_x}\epsilon^{\alpha \beta}H^\beta_z  
\end{equation}
and finally putting all these results in the last equation of (\ref{System 2}), one gets
\begin{equation}
\lambda^3 = \bar{\lambda}_0 z+ \bar{\lambda}_0 \partial_z^{-1} \left(\frac{1+ (H^1_z)^2+ (H^2_z)^2}{H^1_x} \right)
\end{equation}
Note that in the expression of $\lambda^3$ an integration constant that is only dependent on $x$ and $y$ should have appeared, but it must vanish due to the asymptotic behaviour of $\lambda^3$ introduced in (\ref{asym.behaviour of Lagrange mul.}).\\
This ends showing that the solutions of the system of PDEs (\ref{System 2}) are of the form 
\begin{equation}
\Lambda_0^I=\bar{\lambda}_0 \left(x, y,  z+  \partial_z^{-1} \left(\frac{1+ (H^1_z)^2+ (H^2_z)^2}{H^1_x} \right), - \frac{H^2_z}{H^1_x},  \frac{H^1_z}{H^1_x}, 0, \frac{1}{H^1_x} \right)^T
\end{equation}
where $\bar{\lambda}_0 $ is an arbitrary constant.\\
As explained in the paragraph after (\ref{Lambda 1}), it is required to fix $\bar{\lambda}_0 =1$.
Based on the theorem (\ref{th3.2}), to obtain the corresponding physical Hamiltonian for this gauge, $\Lambda_0^I$ should be multiplied to $(h_I)_{G=0}=(0, 0, \partial_I B^I_3, -H^1_x B^y_3, H^1_x B^x_3, 0, 0)$. Consequently, for this special gauge fixing, one gets
\begin{align}\label{Phys. Ham. 4}
h =& \int d^3x\; \left[ \left(z+  \partial_z^{-1} \left(\frac{1+ (H^1_z)^2+ (H^2_z)^2}{H^1_x} \right) \right) \partial_I B^I_3 + \delta^\alpha_I H^\alpha_z B^I_3 \right]\nonumber\\
=&
\int d^3x\; \left[\partial_z^{-1} \left(\frac{1+ (H^1_z)^2+ (H^2_z)^2}{H^1_x}  \right) \partial_I B^I_3 +\epsilon^{IJ} B^I_3 (\partial_z f^3_J) \right] + \oint dS_a (z \delta^a_I B^I_3 )\nonumber\\
=&
\int d^3x\; \left[\partial_z^{-1} \left(\frac{1+ (H^1_z)^2+ (H^2_z)^2}{H^1_x}  \right) \partial_I B^I_3 +\epsilon^{IJ} B^I_3 (\partial_z f^3_J) \right]
\end{align}
Here the surface term has been dropped because according to the boundary conditions (\ref{Asymp B}) and (\ref{New PC1}), it is $O(1)$ odd.\\
\\
The Physical Hamiltonians (\ref{Phys. Ham. 4}) has the following features:\\
I. Linearity in momentum $B^z_\alpha$.\\
II. Non-polynomiality in the configuration variable $f^\alpha_z$.\\
III. Spatial non-locality.\\
\\
In the $(B, f)$ description, comparison of (\ref{Phys. Ham. 4}) and (\ref{Phys. Ham. 3}) tells us that the latter has simpler features, as it is polynomial in the configuration variables. Thus, in the further analysis in future work we will continue with (\ref{Phys. Ham. 3}).

\subsubsection{Remark on polynomial degree of the physical Hamiltonian}

One may ask why it is the case that in some gauges the physical
Hamiltonian
is polynomial while in others it is not. The answer lies in the choice
of the
different polarisations (split
between configuration and momentum degrees of freedom) of the phase
space and
the polynomial degree in which these enter the constraints: In the
(A,E) polarisation, the Gauss constraint is in fact independent of the
momentum $A$
while in the $(B,f)$ polarisation all constraints are linear in the
momentum $B$.
This makes it impossible to impose gauge fixings just in terms of
configuration coordinates $E$ in the $(A,E)$ polarisation while it is
possible for all configuration coordinates $f$ in the $(B,f)$
polarisation.
While $(B,f)$ is a linear canonical transformation of $(A,E)$, it is not
true that the
Gauss constraint div$E=0$ in the $(A,E)$ polarisation and the Bianchi
constraint div$B=0$ in the
$(B,f)$ polarisation are simply rewritings
of each other, in fact they are not at all: Since $B:=\ast dA$ in the
$(A,E)$ polarisation is a derived quantity,
the relation
div$B\equiv 0$ is considered an identity and not as a constraint.
Conversely, since
$E:=\ast df$ is a derived quantity in the $(B,f)$ polarisation, the
relation
div$E\equiv0$ is considered an identity and not as a constraint. On the
other hand
$E$ and $B$ respectively are considered as independent quantities in the
$(A,E)$ and $(B,f)$ polarisation respectively before imposing the
respective Gauss constraints div$E=0$ and div$B=0$. Hence the Gauss and
Bianchi constraint respectively
act on disjoint sets of canonical coordinates.

It transpires that this crucial difference has a major
impact on the available gauge choices and on the entries of the
associated gauge
fixing matrix as far as the polynomial degree with respect to $E$
respectively $f$
is concerned.

\section{Conclusion and Outlook}
\label{s5} 

In this paper, we have provided various indications for the hope 
that the $U(1)^3$ truncation model for Euclidean vacuum quantum gravity 
could have a reduced phase space quantisation with unexpected level of analytical 
control. The essential feature of the model that makes this possible is the fact 
that the constraints are at most linear in momentum (understood as the Abelian 
connection). We have explored various gauges in which the reduced Hamiltonian for 
the true respective degrees of freedom adopts a manageable algebraic form.
We have only sketched the quantisation of the resulting physical degrees 
of freedom and its physical Hamiltonian in this paper but we could already 
argue that the essential feature of the model drastically simplifies the 
spectral problem.
It worth mentioning that concerning the relation between gauge-fixing and gauge-invariant formalisms, one can see that there is a one-to-one correspondence between a choice of gauge fixing and a preferred set of gauge invariant functions which generate the full algebra of gauge invariant functions \cite{17}. The two formalisms are therefore equivalent at generic points of the reduced phase space at which the Dirac matrix (which is a non-trivial function on phase space in every interacting theory) is non-singular. In the same sense, different gauge fixing conditions are generically (i.e., locally in phase space) equivalent. As usual, global differences may have an effect on the quantisation in different gauge choices. However, our attitude is that in quantum gravity global non-equivalence of gauge fixed theories is a second order concern, one would be happy to have at least one working quantisation at one's proposal to start with which then can be further improved. It is in this spirit that the present paper prepares the ground for the reduced phase space quantisation of the $U(1)^3$ test laboratory for LQG.

In the future, we wish to further explore the system in various gauges that may 
simplify the analysis even further and to develop the quantum theory in 
much more detail. Our results could shed new light on the operator 
constraint (or Dirac) approach to the quantisation of the model \cite{7}
in the sense that the physical predictions of the two approaches should agree 
at least semiclassically.\\
\\
{\bf Acknowledgements}\\
\\
S.B. thanks the Ministry of Science, Research and Technology of Iran and 
FAU Erlangen-N\"urnberg for financial support.     

\newpage
\appendix
\section{Generators of asymptotic symmetries in the $(B,f)$ description of the $U(1)^3$ model}\label{App.A}
In this section we examine two boundary conditions. The first one is just transcription of the boundary conditions (\ref{2.1}) (see \cite{11} for more details) while in the other one we use opposite parity conditions. It turns out that the latter is useful to obtain a simple physical Hamiltonian in the process of reduced phase space quantization in the $(B,f)$ description.
\\
\\
A.
{\it Standard parity conditions}\\
\\
As $E^a_i= \delta^a_i + \epsilon^{abc}\partial_b f^i_c$ and $B^a_i= \epsilon^{abc} \partial_b A^i_c$, transcribing the boundary conditions (\ref{2.1}) imposed on $(A, E)$ to the $(B, f)$ description results in
\begin{align}\label{Standard BC}
f_a^i &= c_a^i + \bar{F}_a^i  + O(r^{-1})\nonumber\\
B^a_i &= \frac{\bar{G}^a_i}{r^3}+ O(r^{-4}) 
\end{align}
where $c_a^i$ are constants and $\bar{F}_a^i$ and $\bar{G}^a_i$ are tensor fields defined on the asymptotic 2-sphere with the following definite parity conditions
\begin{equation}\label{Standard PC}
    \begin{split}
        \bar{F}_a^i\left(-\frac{x}{r}\right)= -\bar{F}_a^i\left(\frac{x}{r}\right), \; \; \; \bar{G}^a_i\left(-\frac{x}{r}\right)=\bar{G}^a_i\left(\frac{x}{r}\right).
    \end{split}
\end{equation}
which are direct consequences of the certain parities of the leading terms of $A$ and $E$ (recall that the former is odd and the latter is even).
By the decay conditions (\ref{Standard BC}) and (\ref{Standard PC}), it is assured that the symplectic structure is well-defined.\\

\subsubsection*{Bianchi Constraint}
The Bianchi constraint is 
\begin{equation}
C_i[\Lambda^i]= \int d^3x \; \Lambda^i \partial_a B^a_i
\end{equation}
Since $\partial_a B^a_i$ fall off as $O(r^{-4})$ odd, the minimal
condition on the multiplier $\Lambda^i$ ensuring convergence of the integral is
\begin{equation}\label{BC on lambda}
\Lambda^i = \lambda^i r + O(1)
\end{equation}
where $\lambda^i$ are even functions defined on the asymptotic $S^2$, i.e. 
\begin{equation}
\lambda^i\left(-\frac{x}{r}\right)=\lambda^i\left(\frac{x}{r}\right)
\end{equation}
The action of the Bianchi constraint on the phase space variables is
\begin{equation}
    \begin{split}
        &\delta_{\Lambda}f^j_a=\{C_i[\Lambda^i],f^j_a\}=-\partial_a\Lambda^j\\
        &
        \delta_{\Lambda}B^a_j=\{C_i[\Lambda^i],B^a_j\}=0
    \end{split}
\end{equation}
Thus one sees that
\begin{equation}\label{Cal. for. diff. Bianchi}
    \begin{split}
        \delta C_j[\Lambda^j]=&\int d^3x \; \Lambda^j \partial_a \delta B^a_j=\oint dS_a \; \Lambda^j \delta B^a_j
        -\int d^3x \; (\partial_a\Lambda^j) \delta B^a_j=-\int d^3x \; (\partial_a\Lambda^j) \delta B^a_j\\
        =&
        \int d^3x\; \left[(\delta_\Lambda f^j_a) \delta B^a_j - (\delta_\Lambda B^a_j) \delta f^j_a \right]
    \end{split}
\end{equation}
is functionally differentiable. Here the surface term has been dropped because $\delta B^a_j=O(r^{-3})$ even and $\Lambda^j = O(r)$ even.  
\\

\subsubsection*{Scalar constraint}
It is straightforward to see that
\begin{equation}
    \begin{split}
        &\delta_N f_c^i (x)=\{C[N],f^i_c (x)\}
        =\epsilon_{ikl} \epsilon_{abc} N  (\delta^a_k+ \epsilon^{ade}\partial_d f^k_e) (\delta^b_l+ \epsilon^{bfg}\partial_f f^l_g)\\
        & 
        \delta_N B^c_i (x)=\{C[N],B^c_i (x)\}
        =2\epsilon_{jil} \epsilon_{abe}\epsilon^{adc} \delta  f^i_c \partial_d \left(N  B^e_j  [\delta^b_l+ \epsilon^{bfg}\partial_f f^l_g]\right) 
    \end{split}
\end{equation}
\\
Thus the variation of this constraint is
\begin{align}\label{delta Ham. cons.}
        \delta C[N]=& \int d^3x \; \epsilon_{jkl}N \left(\epsilon_{abc} \delta B^c_j (\delta^a_k+ \epsilon^{ade}\partial_d f^k_e) (\delta^b_l+ \epsilon^{bfg}\partial_f f^l_g)+ 2 \epsilon_{abc} B^c_j (\delta^b_l+ \epsilon^{bfg}\partial_f f^l_g)(\epsilon^{ade}\partial_d \delta  f^k_e) \right)\nonumber\\
        =& \int d^3x \; \epsilon_{jkl} \left(\epsilon_{abc} N \delta B^c_j (\delta^a_k+ \epsilon^{ade}\partial_d f^k_e) (\delta^b_l+ \epsilon^{bfg}\partial_f f^l_g) - 2\epsilon_{abc}\epsilon^{ade} \delta  f^k_e \partial_d \left(N B^c_j  [\delta^b_l+ \epsilon^{bfg}\partial_f f^l_g]\right)  \right)\nonumber\\
        & + 2 \oint dS_d \; \epsilon_{abc}\epsilon^{ade} N  B^c_j  (\delta^b_l+ \epsilon^{bfg}\partial_f f^l_g) \delta  f^k_e \nonumber\\
        =&\int d^3x \; \left(\delta B_j^{a}(\delta_N f_a^j)-(\delta_N B_j^a) \delta f_a^j\right)
\end{align}
where the surface integral vanishes because it is $O(r^{-1})$ even for a translation and $O(1)$ odd for a boost.
Hence, the scalar constraint is functionally differentiable without the need for modification and now its finiteness has to be checked. The Hamiltonian constraint in terms of $(B,f)$ is
\begin{equation}\label{Ham. cons.}
C[N]= \int d^3x \; \epsilon_{jkl}N \left(\epsilon_{abc} B^c_j (\delta^a_k+ \epsilon^{ade}\partial_d f^k_e) (\delta^b_l+ \epsilon^{bfg}\partial_f f^l_g)\right)
\end{equation} 
that is $O(r^{-1})$ even for a translation and $O(1)$ odd for a boost, both of which are divergent. Therefore, neither boosts nor translations have well-defined generators!
\\
\subsubsection*{Vector constraint}
The vector constraint acts on the canonical variables as follows
\begin{equation}
    \begin{split}
        \delta_{\Vec{N}}f^i_c(x)
        =&\{C_a[N^a], f^i_c(x)\} =\epsilon_{abc}N^a(\delta^b_i + \epsilon^{bde} \partial_d f^i_e)\\
        \delta_{\Vec{N}}B^c_i (x) =&\{C_a[N^a], B^c_i (x)\} =\epsilon_{abc} \epsilon^{bde} \partial_d(N^a B^c_i) 
    \end{split}
\end{equation}
So, variation of the constraint is
\begin{equation}\label{variation of diff. constraint}
  \begin{split}
    \delta C_a[N^a]
    &=\int d^3x\;N^a\left(\epsilon_{abc}\delta B^c_j (\delta^b_j + \epsilon^{bde} \partial_d f^j_e) +\epsilon_{abc} \epsilon^{bde} B^c_j \partial_d \delta f^j_e \right)\\
    &=\int d^3x\;\left(\epsilon_{abc}N^a\delta B^c_j (\delta^b_j + \epsilon^{bde} \partial_d f^j_e) -\epsilon_{abc} \epsilon^{bde} \partial_d(N^a B^c_j) \delta f^j_e \right)
     + \oint dS_d \; \epsilon_{abc} \epsilon^{bde} N^a B^c_j \delta f^j_e\\
    =&\int d^3x\;\left(\delta B_j^a\left[\delta_{\vec{N}} f_a^j\right]
    -\delta f_a^j \left[\delta_{\vec{N}} B^a_j \right]\right)
  \end{split}
\end{equation}
Here the surface integral is $O(1)$ odd for a rotation and  $O(r^{-1})$ even for a translation and so can be put away. Thus, the vector constraint 
\begin{equation}\label{Vector const.}
C_a[N^a]
    =\int d^3x\; \epsilon_{abc} N^a B^c_j (\delta^b_j + \epsilon^{bde} \partial_d f^j_e)
\end{equation}
is functionally differentiable but not convergent because it is $O(r^{-1})$ even for a translation and $O(1)$ odd for a rotation. Consequently, $C_a[N^a]$ is not a well-defined generator for translations, nor for rotations!
\\
\\
It is worth mentioning that the source of divergences in (\ref{Ham. cons.}) and (\ref{Vector const.}) can not be written in terms of the constraints, therefore one can not cure the problem. 
\\
\\
B. {\it Boundary conditions with opposite parities}\\
\\
\begin{equation}\label{New PC}
    \begin{split}
        \bar{F}_a^i\left(-\frac{x}{r}\right)= \bar{F}_a^i\left(\frac{x}{r}\right), \; \; \; \bar{G}^a_i\left(-\frac{x}{r}\right)=-\bar{G}^a_i\left(\frac{x}{r}\right).
    \end{split}
\end{equation}
It is obvious that with these new parity conditions still the symplectic structure is well-defined.\\
The Bianchi constraint is 
\begin{equation}
C_i[\Lambda^i]= \int d^3x \; \Lambda^i \partial_a B^a_i
\end{equation}
According to the new boundary conditions $\partial_a B^a_i$ fall off as $O(r^{-4})$ even, thus the minimal
condition on the multiplier $\Lambda^i$ ensuring convergence of Bianchi constraint is the same as (\ref{BC on lambda}) except that $\lambda^i$ has to be an odd function, i.e. 
\begin{equation}
\lambda^i\left(-\frac{x}{r}\right)=-\lambda^i\left(\frac{x}{r}\right)
\end{equation}
In this case, too, $C_j[\Lambda]$ is functionally differentiable because the surface term in (\ref{Cal. for. diff. Bianchi}) vanishes (recall $\delta B^a_j=O(r^{-3})$ odd and $\Lambda^j = O(r)$ odd).
\\
In this case, $C[N]$ is functionally differentiable since the surface integral appearing in (\ref{delta Ham. cons.}) simply vanishes because it is again $O(r^{-1})$ even for a translation and $O(1)$ odd for a boost. But unlike what we observed in the previous case, here by looking at (\ref{Ham. cons.}) one finds that it is $O(r^{-1})$ odd for a translation and $O(1)$ even for a boost. Consequently, translations have a well-defined generator $C[N]$, but boosts do not!  
\\
Again $C_a[N^a]$ is functionally differetiable, since the surface term in (\ref{variation of diff. constraint}) is $O(1)$ odd for a rotation and  $O(r^{-1})$ even for a translation and so it vanishes. It is also convergent for a translation but not for a rotation as (\ref{Vector const.}) is $O(r^{-1})$ odd for a translation and $O(1)$ even for a rotation. Hence, $C_a[N^a]$ is only a well- defined generator for translations but not for rotations. 
\\
\\
The question arising here is why, despite the fact that the boundary conditions are exactly transcribed into the $(B, f)$ description, the use of standard parity conditions leads to well-defined generators in the $(A, E)$ description but not in the $(B,f)$ description. To perceive the source of this discrepancy, first consider a general situation in which we have canonical variables $(A,E)$ and a functional $\mathcal{F}(A,E)$ depending of $A$ only via $\partial_a A$. For simplicity, all the indices of the fields have been dropped.
\\
The variation of $\mathcal{F}[\lambda]$, where $\lambda$ is a test function, would be of the form
\begin{equation}\label{variation in terms of A,E}
\delta \mathcal{F}[\lambda] = \int d^3x  \; \left(\mathcal{A} \partial_a \delta A + \mathcal{B} \delta E \right)= \int d^3x  \; \left(- (\partial_a \mathcal{A}) \delta A + \mathcal{B} \delta E \right) + \oint dS_a \mathcal{A} \delta A
\end{equation}
If one wants to change the canonical variables $(A, E)$ to $(B, f)$ in which $B= \partial_a A$ and $E= c + \partial_a f$ ($c$ is constant), then the variation of $\mathcal{F}$ would be of the form
\begin{equation}\label{variation in terms of B,f}
\delta \mathcal{F}[\lambda] = \int d^3x  \; \left(\mathcal{A}  \delta B + \mathcal{B} \partial_a \delta f \right)= \int d^3x  \; \left(\mathcal{A} \delta B - (\partial_a \mathcal{B}) \delta f \right) + \oint dS_a \mathcal{B} \delta f
\end{equation}
here $\mathcal{A}, \mathcal{B}$ are written in terms of $(B, f)$.
\\
It is obvious from (\ref{variation in terms of A,E}) and (\ref{variation in terms of B,f}) that if we worked with a compact space then $\mathcal{F}$ would be functionally differentiable in terms of both canonical variables. But if one desires to consider boundary conditions, then the story is completely different! Looking at the surface terms in (\ref{variation in terms of A,E}) and (\ref{variation in terms of B,f}), we find that they have nothing in common! Thus, it is quite probable that considering boundary conditions which makes (\ref{variation in terms of A,E}) functionally differentiable keeps (\ref{variation in terms of B,f}) ill-defined and vice versa.
\\
\\
This is exactly what happens when we want to analyse differentiability of the Hamiltonian and diffeomorphism constraints in the $U(1)^3$ model. Recall that the Hamiltonian constraint is $C= 2 \epsilon_{jkl} (\partial_a A^j_b) E^a_k E^b_l$ which depends on $A$ via $\partial_a A$. Therefore,
\begin{align}
\delta C[N] &= \int d^3x \; \left(2N \epsilon_{jkl} (\partial_a \delta A^j_b) E^a_k E^b_l+ 4N \epsilon_{jkl} (\partial_a A^j_b) E^a_k \delta E^b_l  \right)\nonumber\\
&=: \; \int d^3x \; \left( \mathcal{A}^{ab}_j (\partial_a \delta A^j_b) + \mathcal{B}^l_b \delta E^b_l  \right)\label{A4.20}\\
&= \; \int d^3x \; \left( -(\partial_a \mathcal{A}^{ab}_j) \delta A^j_b + \mathcal{B}^l_b \delta E^b_l  \right) + \oint dS_a \mathcal{A}^{ab}_j \delta A^j_b\label{A4.21}
\end{align}
where $\mathcal{A}^{ab}_j:=2N \epsilon_{jkl}E^a_k E^b_l $ and $\mathcal{B}^l_b:=4N \epsilon_{jkl} (\partial_a A^j_b) E^a_k$.
\\
Now, we would like to write $\delta C[N]$ in terms of $(B, f)$ where $B^a_i = \epsilon^{abc} \partial_b A^i_c$ and $E^a_i=\delta^a_i + \epsilon^{abc} \partial_b f^i_c$. We start from (\ref{A4.20}) and write anything in terms of $(B, f)$
\begin{align}
\delta C[N] &= \; \int d^3x \; \left( \mathcal{A}^{ab}_j (\partial_a \delta A^j_b) + \mathcal{B}^l_b \delta E^b_l  \right)\nonumber\\
&= \; \int d^3x \; \left(\frac{1}{2} \mathcal{A}^{ab}_j \epsilon_{cab}  \delta B^c_j + \mathcal{B}^l_b \epsilon^{bac}(\partial_a \delta f_c^l)  \right)\nonumber\\
&= \; \int d^3x \; \left(\frac{1}{2} \mathcal{A}^{ab}_j \epsilon_{cab}  \delta B^c_j - (\partial_a \mathcal{B}^l_b) \epsilon^{bac} \delta f_c^l  \right) + \oint dS_a \; \mathcal{B}^l_b  \epsilon^{bac} \delta f_c^l \label{A4.22}
\end{align}
Now if we use the standard boundary conditions in which $E^a_i - \delta^a_i = O(r^{-1})$even, $A^i_a = O(r^{-2})$odd and consequently $B^a_i= O(r^{-3})$even, $f^i_a=O(1)$odd, then it is concluded that for translations $\mathcal{A}^{ab}_j=\text{constant}+O(r^{-1})$even and $\mathcal{B}^l_b = O(r^{-3})$even.
\\
Hence, the surface terms in (\ref{A4.21}) and (\ref{A4.22}) are 
\begin{align}
&\oint dS_a \mathcal{A}^{ab}_j \delta A^j_b= \oint dS_a \; O(r^{-2})\text{odd} = \text{divergent}\\
&\oint dS_a \; \mathcal{B}^l_b  \epsilon^{bac} \delta f_c^l = \oint dS_a \; O(r^{-3})\text{odd}=0
\end{align}
Therefore, $C[N]$ is functionally differentiable in the $(B,f)$ description but not in the $(A,E)$ description!
\\
Hence, to get a well-defined generator for the asymmetric temporal translation, we have to add a term to the Hamiltonian constraint  eliminating the divergence appearing in its variation. In this way, one obtains an expression which is
functionally differentiable and, as one is lucky, is already finite (for more details see \cite{11}).
On the other hand, in the $(B,f)$ description, because there does not exist such a divergence in the variation of $C[N]$, it is already functionally differentiable, and the only factor making the Hamiltonian constraint ill-defined is its own divergence. Since the origin of this divergence cannot be written in terms of the constraints, it is not admissible to subtract it from the Hamiltonian constraint, therefore $C[N]$ remains ill-defined in the $(B,f)$ description. 
The same happens for the diffeomorphism constraint.

\section{Solutions of the constraints in the $(B,f)$ description}\label{App. B}
Due to the aforementioned reason in theorem (\ref{th3.2}), the solutions of the constraints are not required to attain the physical Hamiltonian. However, to ensure that the model is consistent, we must answer the question of whether the system of equations under consideration admits solutions satisfying the asymptotic behaviour determined. For this purpose, in this appendix, we exhibit that by considering the boundary conditions (\ref{Asymp B}), (\ref{Asymp f}) and (\ref{New PC}), there exist solutions for the systems of equations (\ref{B MTA-MLA}) and (\ref{B ETTA-ELA}).

\subsection{Solutions for the system of equations (\ref{B MTA-MLA})}
Inspecting the system of equations (\ref{B MTA-MLA}), one can solve the first and second equations for $B^x_\alpha$ as $B^x_\alpha= -\partial^{-1}_x (\partial_y B^y_\alpha + \partial_z B^z_\alpha)+ g_\alpha (y,z)$ where $g_\alpha$ are arbitrary functions depending only on $y,z$. As $B^x_\alpha$ decays at infinity due to (\ref{Asymp B}), $g_\alpha$ have to vanish. Therefore,
\begin{equation}\label{B.1}
B^x_\alpha= -\partial^{-1}_x (\partial_y B^y_\alpha + \partial_z B^z_\alpha).
\end{equation} 
The third equation of (\ref{B MTA-MLA}) is solved for $B^z_3$ as $B^z_3= -\partial_z^{-1}\partial_I B^I_3 + g(x,y)$ in which $g$ is an arbitrary function in the kernel of $\partial_z$. Again, using the asymptotic behaviour of $B^z_3$, one concludes $g=0$. Hence,
\begin{equation}\label{B.2}
B^z_3= -\partial_z^{-1}\partial_I B^I_3
\end{equation}
Since $H^3_z = \det (H^\alpha_I)\neq 0$, the fourth and fifth equations of (\ref{B MTA-MLA}) that simply form an algebraic system of two equations with two unknowns $B^I_3$ can be solved as
\begin{equation}\label{B.3}
B^I_3= \epsilon^{IJ} \epsilon^{\alpha \beta} B^z_\alpha H^\beta_J
\end{equation}
One plugs (\ref{B.1})-(\ref{B.3}) into the sixth and seventh equations of (\ref{B MTA-MLA}) and gets
\begin{align}
& - Y_2 B^y_1 + Y_1 B^y_2 = \epsilon^{\alpha \beta} H^\beta_x \partial_x^{-1} \partial_z B^z_\alpha \label{B.4}\\
& - Y_1 B^y_1 - Y_2 B^y_2 = H^\alpha_x \partial_x^{-1}\partial_z B^z_\alpha + \epsilon^{IJ} \epsilon^{\alpha \beta} H^3_z \partial_z^{-1} \partial_I (H^\beta_J B^z_\alpha) \label{B.5}
\end{align}
where $Y_\alpha := H^\alpha_x \partial_x^{-1} \partial_y - H^\alpha_y$. Using the inverse of the operator $Y_2$, we can solve (\ref{B.4}) for $B^y_1$ as $  B^y_1 = Y_2^{-1} Y_1 B^y_2 - Y_2^{-1} \left(\epsilon^{\alpha \beta} H^\beta_x \partial_x^{-1} \partial_z B^z_\alpha\right)+ \kappa$ in which $\kappa$ is in the kernel of $Y_2$ and of the form $\kappa= \sum_{n=3}^\infty \kappa_n r^{-n}$, since $B^y_1 = O(r^{-3})$. As $H^\alpha_I = \delta^\alpha_I + O(r^{-1})$, the highest order term of the defining equation for $\kappa$, that is $0 = Y_2 \kappa = H^2_x \partial_x^{-1} \partial_y \kappa - H^2_y \kappa$, is $\kappa_3 / r^{3}$ which has to vanish individually. Hence, $\kappa_3 = 0$ and $\kappa= O(r^{-4})$. Repeating the same argument, one deduces that $\kappa = 0$. Therefore, 
\begin{equation}\label{B.6}
B^y_1 = Y_2^{-1} Y_1 B^y_2 - Y_2^{-1} \left(\epsilon^{\alpha \beta} H^\beta_x \partial_x^{-1} \partial_z B^z_\alpha\right)
\end{equation}
Inserting (\ref{B.6}) in (\ref{B.5}) leads to
\begin{equation}\label{B.7}
X B^y_2 = Y_1 Y_2^{-1} \left(\epsilon^{\alpha \beta} H^\beta_x \partial_x^{-1} \partial_z B^z_\alpha\right) -H^\alpha_x \partial_x^{-1}\partial_z B^z_\alpha - \epsilon^{IJ} \epsilon^{\alpha \beta} H^3_z \partial_z^{-1} \partial_I (H^\beta_J B^z_\alpha) 
\end{equation}
where $X := Y_1 Y_2^{-1} Y_1 + Y_2$.
Using the inverse of $X$, we solve (\ref{B.7}) for $B^y_2$ as $B^y_2 =X^{-1}  Y_1 Y_2^{-1} \left(\epsilon^{\alpha \beta} H^\beta_x \partial_x^{-1} \partial_z B^z_\alpha\right) - X^{-1} \left(H^\alpha_x \partial_x^{-1}\partial_z B^z_\alpha\right) - \epsilon^{IJ} \epsilon^{\alpha \beta} X^{-1} \left( H^3_z \partial_z^{-1} \partial_I (H^\beta_J B^z_\alpha)\right)+ \bar{\kappa} $, in which $\bar{\kappa}$ is a member of the kernel of $X$ and decays at infinity as $O(r^{-3})$, because $B^y_2 = O(r^{-3})$.
In the following, we will use the same method as described in full detail in the subsection (4.2.3) in order to specify $\bar{\kappa}$. Note that the decaying behaviour of $H^\alpha_I$ tells us that $Y_1 = \partial_x^{-1}\partial_y + O(r^{-1})$. Moreover, if one makes use of (\ref{inverse of sum}), it is easily concluded that $Y_2^{-1}= -1 + O(r^{-1})$. Thus, the highest order term of the r.h.s. of the defining equation for $\bar{\kappa}$, i.e. $0= X \bar{\kappa}= (Y_1 Y_2^{-1} Y_1 + Y_2)\bar{\kappa} $, is $-(\partial_x^{-2}\partial_y^2 + 1) \frac{\bar{\kappa}_3}{r^3}$ that has to vanish separately. Applying $\partial_x^2$, we see that the highest order term of $\bar{\kappa}$ must satisfy the 2-dimensional Laplace equation $\Delta (\bar{\kappa}_3/r^3)=0$ which means that $\bar{\kappa}_3 = 0$, since in $\mathbb{R}^3$ the only  harmonic function decaying at infinity is the trivial function. Therefore, $\bar{\kappa}=O(r^{-4})$. Iterating the argument results in $\bar{\kappa}=0$. So,
\begin{equation}\label{B.8}
B^y_2 =X^{-1}  Y_1 Y_2^{-1} \left(\epsilon^{\alpha \beta} H^\beta_x \partial_x^{-1} \partial_z B^z_\alpha\right) - X^{-1} \left(H^\alpha_x \partial_x^{-1}\partial_z B^z_\alpha\right) - \epsilon^{IJ} \epsilon^{\alpha \beta} X^{-1} \left( H^3_z \partial_z^{-1} \partial_I (H^\beta_J B^z_\alpha)\right)
\end{equation}
This ends showing that by solving the constraints, $B^I_i$ and $B^z_3$ can be expressed in terms of $B^z_\alpha$ which are our degrees of freedom.
\subsection{Solutions for the system of equations (\ref{B ETTA-ELA})}
Looking at the system of equations (\ref{B ETTA-ELA}), we can solve the first and second equations for $B^x_\alpha$ as $B^x_\alpha= -\partial^{-1}_x (\partial_y B^y_\alpha + \partial_z B^z_\alpha)+ g_\alpha (y,z)$ where $g_\alpha$ are arbitrary functions in the kernel of $\partial_x$. Due to the decaying behaviour (\ref{Asymp B}), $g_\alpha$ have to vanish. Therefore,
\begin{equation}\label{B.9}
B^x_\alpha= -\partial^{-1}_x (\partial_y B^y_\alpha + \partial_z B^z_\alpha).
\end{equation} 
One solves the third equation of (\ref{B ETTA-ELA}) for $B^z_3$ as $B^z_3= -\partial_z^{-1}\partial_I B^I_3 + g(x,y)$ in which $g$ is an arbitrary function depending only on $x, y$. Again, $g=0$ follows from the asymptotic behaviour of $B^z_3$. Thus,
\begin{equation}\label{B.10}
B^z_3= -\partial_z^{-1}\partial_I B^I_3
\end{equation}
Solving the fourth and fifth equations of (\ref{B ETTA-ELA}) for $B^z_\alpha$ simply leads to
\begin{equation}\label{B.11}
B^z_\alpha = H^1_x \delta^\alpha_I B^I_3 - H^\alpha_z \partial_z^{-1} \partial_I B^I _3
\end{equation}
Since $H^1_x \neq 0$, the last two equations of (\ref{B ETTA-ELA}) can be rewritten as
\begin{align}
&B^y_1 = B^x_2 - \frac{1}{H^1_x} \epsilon^{\alpha \beta} B^z_\alpha H^\beta_z \label{B.12}\\
&B^x_1 + B^y_2 = - \frac{1}{H^1_x} \left(H^\alpha_z B^z_\alpha + B^z_3 \right)\label{B.13}
\end{align}
If one employs (\ref{B.9})-(\ref{B.12}) in (\ref{B.13}) and simplifies, the following integro-differential equation arises
\begin{equation}\label{B.14}
\partial^{-2}_x \Delta B^y_2 = - \frac{1}{H^1_x} \left(H^\alpha_z B^z_\alpha + B^z_3 \right) - \partial^{-1}_x \partial_y\left(\frac{1}{H^1_x} \epsilon^{\alpha \beta} B^z_\alpha H^\beta_z \right) +\partial^{-1}_x \partial_z B^z_1 - \partial^{-2}_x \partial_y  \partial_z B^z_2 
\end{equation}
which can be solved for $B^y_2$ using the inverse of $\partial^{-2}_x \Delta$. We denote the r.h.s. of (\ref{B.14}) which depends only on $B^I_3$ by $R(B^I_3)$ and see that $B^y_2 = \Delta^{-1} \partial_x^2 R + \kappa$ where $\partial^{-2}_x \Delta \kappa = 0$ and $\kappa = O(r^{-3})$ due to (\ref{Asymp B}). The asymptotic behaviour of $\kappa$ shows that $\Delta \kappa$ has to vanish and because in $\mathbb{R}^3$ the only harmonic function approaching zero at infinity is the trivial function, $\kappa=0$. Therefore,
\begin{equation}
 B^y_2 =  \Delta^{-1}\left[ -\partial^2_x \left( \frac{1}{H^1_x} \left(H^\alpha_z B^z_\alpha + B^z_3 \right)\right) - \partial_x \partial_y\left(\frac{1}{H^1_x} \epsilon^{\alpha \beta} B^z_\alpha H^\beta_z \right) +\partial_x \partial_z B^z_1 - \partial_y  \partial_z B^z_2 \right]
\end{equation} 
Hence, we exhibited that by solving the constraints, $B^I_\alpha$ and $B^z_i$ can be expressed in terms of $B^I_3$ which are our degrees of freedom.

\end{document}